\documentclass[letter,aps,10pt,amsmath,showpacs,amssymb,prd,onecolumn,floatfix,nofootinbib]{revtex4}

\usepackage{graphicx}
\usepackage{color,pstcol} 
\usepackage{afterpage}
\usepackage{amsmath}
\usepackage{subfigure}
\usepackage{comment}
\usepackage{lgrind}
\usepackage{dcolumn}
\usepackage{bm}
\usepackage{epsfig} 
\usepackage{natbib}

\def\be{\begin{eqnarray}} \def\ee{\end{eqnarray}}
\def\ba{\begin{eqnarray}} \def\ea{\end{eqnarray}}

\begin{document}

\author{Jamie  Portsmouth}  
\affiliation{Astrophysics, Department of Physics, Keble Road, Oxford,
  OX1 3RH, UK}
\email{jamiep@astro.ox.ac.uk}

\title{Analysis of the Kamionkowski-Loeb method of reducing \\
cosmic variance with CMB polarizationls}

\begin{abstract}
  Part of the CMB polarization signal in the direction of galaxy
  clusters is produced by Thomson scattering of the CMB temperature
  quadrupole. In principle this allows measurement of 
  the CMB power spectrum harmonic $C_2(z)$ with higher accuracy
(at $z>0$) than the cosmic variance limit imposed by sample variance
  on one CMB sky.
  However the observed signals are statistically correlated if
  the comoving separation between the clusters is small enough.
  Thus one cannot reduce the sample variance by more than roughly the
  number of separate regions available which produce uncorrelated
  signals, as first pointed out by Kamionkowski and Loeb.
  In this paper we analyze in detail the procedure outlined by
  Kamionkowski and Loeb,  computing the correlation of the
  polarization signals by considering the variation of the spherical
  harmonic expansion coefficients of the temperature anisotropy on our
  past light cone.

  Given a hypothetical set of Stokes parameter measurements of the CMB
  polarization in the directions of galaxy clusters, distributed at
  random on a given redshift shell, we show how to
  construct an estimator of the angular power spectrum 
  harmonic $C_2$ at that redshift. We then compare the variance of
  this estimator with the cosmic variance of
  the CMB multipole on our sky which probes the same scale.
  We find that in fact the cosmic variance is not reducible
  below the single sky CMB value using the cluster method.
  Thus this method is not likely to be of use for reconstruction
  of the primordial power spectrum. However the method does yield a
  measurement of $C_2$ as a function of redshift with increasing
  accuracy at higher redshift, and thus potentially a probe of the
  mechanism which may have suppressed the quadrupole.

  We also examine to what extent the redshift dependence of $C_2$ can
  be used to probe the time changing potential anisotropy as the
  universe evolves into the vacuum dominated phase (the late-time
  integrated Sachs-Wolfe  effect). We find that this effect is not
  observable in the time dependence of $C_2$ since it is swamped by
  cosmic variance, but there is an observable signature in the correlation
  functions of the Stokes parameters.

\end{abstract}
\pacs{98.80.Es,95.30.Gv,98.70.Vc}

\date{\today}

\maketitle

\setcounter{section}{0}

\section{Introduction}

The CMB radiation incident on galaxy clusters has an intrinsic
intensity quadrupole $Q_2$ created
by inhomogeneity at the surface of last scattering. 
Thomson scattering of the CMB in a galaxy cluster with typical
line of sight optical depth $\tau_{\rm C}$ generates
polarization of order $Q_2 \tau_{\rm C}$.  
Thus a measurement of this polarization
signal would allow an estimate of the CMB quadrupole at non-zero
redshift. 
This is of interest because it would potentially allow
us to get around the restriction of cosmic variance.
To elaborate, at $z=0$ we only have one CMB sky to observe,
with $(2l+1)$ independent real data points for
each spherical harmonic mode of the CMB on our sky, to compare with the ensemble average prediction of the variance.
There is thus an intrinsic fractional sample variance of the harmonic
$C_l$ of $2/(2l+1)$ (see section \S\ref{ch5:gencorr}), 
which severely limits comparison with the ensemble
averaged theory at low $l$. This restriction limits the accuracy of
measurements of the primordial power spectrum on
the largest scales. The theoretical predictions thus obtained for
the CMB power spectra are fundamentally limited by this sample
variance, commonly termed the \emph{cosmic variance}.

Thomson scattering of the $l=2$ part of the CMB anisotropy
in a cluster generates a secondary polarization anisotropy which
depends on the spherical harmonic components $a_{2m}$ as seen by a
(hypothetical) observer at the cluster.
Since this polarization signal produced by a cluster is sensitive to
the density perturbations on a last scattering surface different to
our own, this in principle allows one to make more accurate
comparison to the theoretical predictions for CMB angular power
spectra at low $l$ than allowed by the cosmic variance limit. 
However the observed signals are correlated if
the comoving separation between the clusters is small,
and many strongly correlated signals are no more useful
for reducing the sample variance than one signal.
The variance in the estimated quadrupole can be
reduced by roughly the number of regions available which produce
uncorrelated signals. 

This method of using the CMB polarization signal produced by galaxy
clusters to get around cosmic variance limits was first pointed out by
Kamionkowski and Loeb \cite{Kamionkowski:1997na}
(we usually refer to it as the ``cluster method'' in what follows).
However Kamionkowski and Loeb did not
actually compute the correlation of the cluster signals in a
particular cosmological model in their paper, and did not therefore
demonstrate explicitly that the cosmic variance is reduced with a
given set of clusters, nor did they develop any formalism for converting
measurements of the Stokes parameters of the CMB to 
statistical estimators which get around cosmic variance.
In \cite{2003PhRvD..67f3505C}, estimators of
$C_2(\tau)$ were constructed
(taking into account the kinematic SZ contamination of the
polarization signal also), but the effect of
statistical variation in the polarization signal on the estimator
variance was not included.
This variation was considered by \cite{2000PhRvD..62l3004S}, 
but they computed the variation of the quadrupole as
an expansion in small cluster separations, and their
analysis is not applicable to a general set of clusters
in arbitrary locations.
In this paper we compute the correlation of the cluster signals in
the case of an idealized set of measurements from clusters distributed
in random directions on a given redshift shell. 
We describe an explicit procedure for carrying out the program
outlined in  \cite{Kamionkowski:1997na}, and study how the
correlations die off  as clusters of increasingly high redshift are
used. 
Information about the correlation of the polarization signals is
contained in the generalized correlation functions of the CMB
temperature anisotropy coefficients, 
$\langle a_{lm}(\mbox{\boldmath $x$},\tau)
a^*_{l^{\prime}m^{\prime}}(\mbox{\boldmath
  $x$}^{\prime},\tau^{\prime}) \rangle$,
which contain all of the statistical information 
(assuming Gaussianity) about the variation of the $a_{lm}$
coefficients as the observation point and associated last scattering surface change. 
With these functions, we can derive an estimator for $C_2(z)$ 
in terms of Stokes parameters, and find its variance.

We should clarify exactly what we mean by reducing
cosmic variance. It is true that the polarization signals provide an
estimator \( \hat C_2(\tau) \) of the remote quadrupole at a given
redshift which has a smaller fractional cosmic variance than the
local quadrupole.  
However, this is not the most useful comparison, since this
estimator probes smaller physical scales than the local quadrupole.
Cosmologists already have estimators of the power on these scales,
namely the WMAP angular power spectrum harmonics $C_l$ with $l>2$.
So the interesting question to ask is whether \( \hat C_2(\tau) \) has
smaller cosmic variance than \emph{the CMB multipole on our sky 
that probes the same physical scale}. 
This determines whether or not the cluster
technique is capable in principle of providing a better reconstruction
of the primordial potential than the WMAP data.
The results are presented in \S\ref{ch5:sec_QUstat}.

We now outline the organization of the paper.
In \S\ref{ch5:gencorr} 
we derive the two-point generalized correlation
functions of the spherical harmonic coefficients, assuming a Gaussian
primordial perturbation spectrum.
In \S\ref{ch5:sec_transfer} we discuss the CMB transfer
functions used to compute the generalized correlation functions,
and examine the time dependence of $C_2(\tau)$.
In \S\ref{ch5:quadscatt} we derive expressions for 
the Stokes parameters $Q, U$ (defined in an appropriate all-sky basis)
of the CMB radiation scattered into the line-of-sight by the
cluster gas, in terms of the local $a_{lm}$ at the cluster, for a
general line-of-sight. Note that in this section we found
it convenient to use the ``density matrix'' formalism
for polarization calculations
\citep{1994PhDT........24K,2000PhRvD..62d3004C}, which is outlined in
Appendix \ref{appb}.  
Then in \S\ref{ch5:sec_QUstat} we consider the 
the statistical variation of these Stokes parameters 
with the comoving position of the cluster.
We construct a simple estimator for 
$C_2(\tau)$ and compute its variance for a number
of simulated sets of clusters.
In \S\ref{ch5:sec_discuss} there is a discussion and summary.
Note that we restrict the discussion to the case of a flat FRW
universe throughout, for simplicity.

\section{Generalized correlation functions \label{ch5:gencorr}}

The observed CMB sky and associated power spectrum changes
with the comoving spatial position and redshift of the observer.
The statistical variation with position may be characterized
completely with a set of correlation functions of the temperature
perturbation.

The fractional CMB temperature perturbation $\Delta(\mbox{\boldmath
  $x$},\mbox{\boldmath $\hat{n}$},\tau)=\Delta T/T_{\rm CMB}$ 
is a function of the comoving spatial position of the observer
$\mbox{\boldmath $x$}$, the conformal time of observation $\tau$, and
the direction of the line of sight $\mbox{\boldmath $\hat{n}$}$.
In order to expand the CMB temperature field in
spherical harmonics $a_{lm}(\mbox{\boldmath   $x$},\tau)$ we must
define a polar coordinate system everywhere.  In flat space, it is
simplest to use the convention that the polar axis is taken to be the
same at each point (in a curved space, this would not be possible and
more care would be needed).

When we need to specify coordinates for $\mbox{\boldmath $\hat{n}$}$
we use spherical polars $(\theta,\phi)$, with the polar axis
 aligned with the $\mbox{\boldmath $e$}_z$ direction and the 
$\phi=0$ plane normal to the $\mbox{\boldmath $e$}_y$ direction.  
The fractional temperature perturbation is expanded as
\begin{eqnarray} \label{ch4:temp}
&&\Delta(\mbox{\boldmath $x$},\mbox{\boldmath $\hat{n}$},\tau) 
= \sum_{l=0}^{\infty}\sum_{m=-l}^{l}
a_{lm}(\mbox{\boldmath $x$},\tau) Y_{lm}(\mbox{\boldmath $\hat{n}$}) \ .
\end{eqnarray}
The two point correlation function of the temperature anisotropy is
\begin{eqnarray} \label{ch4:clm}
&&\langle \Delta(\mbox{\boldmath $x$},\mbox{\boldmath $\hat{n}$},\tau) 
\Delta(\mbox{\boldmath $x$}^{\prime},\mbox{\boldmath
  $\hat{n}$}^{\prime},\tau^{\prime}) \rangle = 
\sum_{lm} \sum_{l^{\prime} m^{\prime}}
C_{l m l^{\prime} m^{\prime}}(\mbox{\boldmath $x$},\tau,
\mbox{\boldmath $x$}^{\prime},\tau^{\prime})
Y_{lm}(\mbox{\boldmath $\hat{n}$})
Y^*_{l^{\prime}m^{\prime}}(\mbox{\boldmath $\hat{n}$}^{\prime}) \ ,
\quad\quad
\end{eqnarray}
where we have defined the generalized CMB correlation functions
\begin{eqnarray}
C_{l m l^{\prime} m^{\prime}}(\mbox{\boldmath $x$},\tau,
\mbox{\boldmath $x$}^{\prime},\tau^{\prime})
\equiv \langle a_{lm}(\mbox{\boldmath $x$},\tau)
a^*_{l^{\prime}m^{\prime}}(\mbox{\boldmath
  $x$}^{\prime},\tau^{\prime}) \rangle  \ .
\end{eqnarray}

The set of functions $C_{l m
  l^{\prime} m^{\prime}}(\mbox{\boldmath $x$},\tau, \mbox{\boldmath
  $x$}^{\prime},\tau^{\prime})$ form the covariance function of the
Gaussian random process from which the photon distribution function,
defined at all points in space and observed in all directions, is
sampled. 

Given a set of cosmological parameters, the generalized correlation
functions may be computed using the photon transfer function which
describes the physics of the propagation of CMB photons from the last
scattering surface to the cluster.
The generalized correlation functions manifestly obey some simple
symmetry relations:
\ba \label{ch4:symm1}
C_{l,-m,l^{\prime},-m^{\prime}}(\bm{x},\bm{x}^{\prime}) &=&
(-1)^{m+m^{\prime}}C^*_{lml^{\prime}m^{\prime}}(\bm{x},\bm{x}^{\prime})
\ ,
\nonumber \\ 
C_{lml^{\prime}m^{\prime}}(\bm{x}^{\prime},\bm{x}) &=&
C^*_{l^{\prime}m^{\prime}lm}(\bm{x},\bm{x}^{\prime}) \ . 
\ea

We will be interested only in the case where both sets of spacetime
coordinates lie on our past light cone, in order that we are computing
only quantities which are directly observable. 
Choosing ourselves to be at the spatial origin of the comoving
coordinate system, at conformal time $\tau_0$ (the age of the universe
in conformal time), the events at $\mbox{\boldmath $x$}$,
$\mbox{\boldmath $x$}^{\prime}$ occurred at conformal times
$\tau=\tau_0-\vert\mbox{\boldmath $x$}\vert$, $\tau^{\prime}=\tau_0-\vert\mbox{\boldmath $x$}^{\prime}\vert$ respectively.
The correlation function may therefore be written as a function of
spatial variables only, $C_{l m l^{\prime} m^{\prime}}(\mbox{\boldmath
  $x$}, \mbox{\boldmath $x$}^{\prime})$.
The coordinate system is illustrated in Figure \ref{coord
  system}. Note that the last scattering surfaces of observers at
$z>0$ are spheres which are tangential to the last scattering sphere
of an observer at $z=0$.

\begin{figure}
\begin{center}
  \includegraphics[width=7cm]{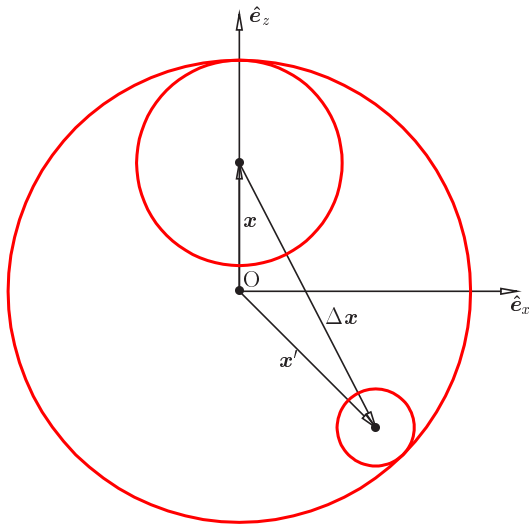}
\caption{Comoving coordinate system for generalized CMB correlation
  function, in flat space. All points in the plane shown lie on the
  past light cone of the observer at O.  The outermost
  circle indicates our last scattering surface. The
  circles centered on the observers at positions $\bm{x}$ and
  $\bm{x}^{\prime}$ (and conformal times 
  $\tau = \tau_0 - \vert\bm{x}\vert$ and $\tau^{\prime} = \tau_0 -
  \vert\bm{x}\vert$ respectively) indicate their last scattering surfaces, which are
  smaller since recombination occurred in the less distant past
  according to them.  Note that any orientation of the points $\bm{x}$
  and $\bm{x}^{\prime}$ in space can be rotated into this plane.  
\label{coord system}
}
\end{center}
\end{figure}

Now we briefly discuss the statistical properties of the coefficients
$a_{lm}$ and review the usual definition of cosmic variance.  
The $a_{lm}$ coefficients at any given point are all
independent, but there are spatial and temporal correlations between
any pair of coefficients at different points.  
Assuming Gaussianity of the primordial perturbations, it 
follows that all of the various $N$-point joint probability
distribution functions for the $a_{lm}$ at separate points 
are (complex) multivariate Gaussians, with covariance
matrix $\bm{R}$ given by (note that each index
labels both the set of values $l_i, m_i$ and also the point
$\mbox{\boldmath $x$}_i$ in three dimensional space):
\begin{eqnarray}
R_{ij} = \langle a_{l_i m_i}(\mbox{\boldmath $x$}_i) a^*_{l_j
m_j}(\mbox{\boldmath $x$}_j)\rangle = C_{l_i m_i l_j
m_j}(\mbox{\boldmath $x$}_i,\mbox{\boldmath $x$}_j) \ .
\end{eqnarray}
Thus given a cosmological model we have the p.d.f of the ensemble from
which the $a_{lm}(\mbox{\boldmath $x$})$ are drawn, and the associated
ensemble average angular power spectrum harmonics $C_l(\tau)$.  Then
given a set of observations $a^o_{lm}(\mbox{\boldmath $x$})$, these
ensemble average predictions are typically compared to the observed
quantities $C^o_l(\mbox{\boldmath $x$})=\sum_{m=-l}^l \vert
a^o_{lm}(\mbox{\boldmath $x$})\vert ^2/(2l+1)$ (clearly $\langle
C_l^o(\mbox{\boldmath $x$})\rangle =C_l(\tau)$). 

On the sky of an observer at time $\tau$, the mean square difference
between the observed CMB angular power spectrum $C_l^o(\bm{x})$ and
the ensemble average theoretical power spectrum $C_l(\tau)$ is
characterized by the cosmic variance 
\begin{eqnarray}
\langle [C_l^o(\mbox{\boldmath $x$})-C_l(\tau)]^2\rangle 
= \langle C_l^o(\mbox{\boldmath $x$})^2\rangle -C_l(\tau)^2 \ .
\end{eqnarray}
Expanding we obtain
\begin{eqnarray}
\langle C_l^o(\mbox{\boldmath $x$})^2\rangle 
&=& \sum_{mm^{\prime}} (2l+1)^{-2} \langle 
\vert a^o_{lm}(\mbox{\boldmath $x$})\vert ^2 \vert a^o_{lm^{\prime}}(\mbox{\boldmath
$x$})\vert ^2 \rangle  \ .
 \end{eqnarray}
To evaluate the right hand side, we need the ensemble
average of the product of four $a_{lm}$'s. 
This follows from Gaussianity:
\begin{eqnarray}\label{ch5:4point}
&&
\!\!\!\!\!\!\!\!\!\!\!\!\!\!\!\!
\langle a_{lm_1}(\mbox{\boldmath $x$},\tau)a^*_{lm_2}(\mbox{\boldmath
    $x$},\tau) a_{l^{\prime}m_3}(\mbox{\boldmath
    $x$}^{\prime},\tau^{\prime})a^*_{l^{\prime}m_4}(\mbox{\boldmath
    $x$}^{\prime},\tau^{\prime}) \rangle  
= 
\delta_{m_1 m_2} \delta_{m_3 m_4} C_l(\tau)
    \;C_{l^{\prime}}(\tau^{\prime}) 
\nonumber \\
&& 
+ \;C_{lm_1 l^{\prime}m_4}(\mbox{\boldmath $x$},\mbox{\boldmath
    $x$}^{\prime}) C^*_{lm_2 l^{\prime}m_3}(\mbox{\boldmath $x$},\mbox{\boldmath $x$}^{\prime}) 
+ C_{lm_1 l^{\prime},-m_3}(\mbox{\boldmath $x$},\mbox{\boldmath
    $x$}^{\prime}) C^*_{lm_2 l^{\prime},-m_4}(\mbox{\boldmath
    $x$},\mbox{\boldmath $x$}^{\prime})  \ . 
\end{eqnarray}
Now setting $l^{\prime}=l$, $\mbox{\boldmath $x$} = \mbox{\boldmath
  $x$}^{\prime}$, and using
$C_{lml^{\prime}m^{\prime}}(\mbox{\boldmath $x$},\mbox{\boldmath $x$})
= \delta_{ll^\prime}\delta_{mm^{\prime}} C_l(\tau)$, we obtain the
  familiar expression for the cosmic variance associated with each
  harmonic, 
\begin{eqnarray} \label{ch5:cvar}
\langle C_l^o(\mbox{\boldmath $x$})^2\rangle  - C_l(\tau)^2
&=& \frac{2}{2l+1} \;C_l(\tau)^2 \ .
\end{eqnarray}
This quantity captures how much we can expect the measured power
spectra to differ from the ensemble average. Note that this expression
also follows from the fact that $C_l^o$ is the sum of squares of
independent identical Gaussian random variables, and is therefore
distributed as a scaled $\chi^2_{2l+1}$ random variable.

Related results concerning the spatial correlations of various
quantities can be derived similarly.
The ensemble average difference between the $a_{lm}$ measured
by separated observers on our past light cone, for example, is
characterized by 
\begin{eqnarray}
\langle \left|a_{l'm'}(\mbox{\boldmath $x'$})-a_{lm}(\mbox{\boldmath
    $x$})\right|^2\rangle &=& C_{l'}(\tau') + C_{l}(\tau) - 2 \mbox{Re}\;
    C_{lml'm'}(\bm{x},\bm{x}') \ ,
\end{eqnarray}
and the ensemble average difference between the angular power spectrum
harmonics as a function of spatial separation is characterized by
\begin{eqnarray}
\left<[C^o_{l}(\mbox{\boldmath $x$})-C^o_{l^{\prime}}(\mbox{\boldmath
$x$}^{\prime})]^2\right> &=& \left<C_l^o(\mbox{\boldmath $x$})^2\right> + 
\left<C_{l^{\prime}}^o(\mbox{\boldmath $x$}^{\prime})^2\right>
- 2 \left<C^o_{l}(\mbox{\boldmath $x$})
C^o_{l^{\prime}}(\mbox{\boldmath $x$}^{\prime})\right> \nonumber \\ 
&=& \left[C_l(\tau) - C_{l^{\prime}}(\tau^{\prime})\right]^2 
+ \frac{2}{2l+1} C_l(\tau)^2 + \frac{2}{2l'+1}
C_{l^{\prime}}(\tau^{\prime})^2 \nonumber \\ 
&-& \frac{2}{(2l+1)(2l^{\prime}+1)} \sum_{mm^{\prime}} \left[
|C_{lml^{\prime}m^{\prime}}(\mbox{\boldmath $x$},\mbox{\boldmath
$x$}^{\prime})|^2 + |C_{lml^{\prime},-m^{\prime}}(\mbox{\boldmath $x$},\mbox{\boldmath
$x$}^{\prime})|^2 \right] \ .
\end{eqnarray}
 
We now derive the relationship between the generalized correlation functions and
the CMB transfer function.  In a flat FRW space, the temperature
anisotropy may be Fourier expanded in comoving wavenumber $\bm{k}$ on
the three-dimensional hyper-surface of constant $\tau$, $\Sigma_\tau$, 
\begin{eqnarray}
\Delta(\mbox{\boldmath $x$},\mbox{\boldmath $\hat{n}$},\tau) = \int
d^3k \;e^{i\mbox{\boldmath $k\cdot x$}} \Delta(\mbox{\boldmath
$k$},\mbox{\boldmath $\hat{n}$};\Sigma_{\tau}) \ ,
\end{eqnarray}
where $\Delta(\mbox{\boldmath $k$},\mbox{\boldmath
  $\hat{n}$};\Sigma_{\tau})$ is the Fourier transform associated with
this hyper-surface only \citep{1995ApJ...455....7M}.  

Since each Fourier mode
corresponds in the case of
scalar perturbations to a plane wave perturbation which has azimuthal
symmetry about $\mbox{\boldmath $k$}$, $\Delta(\mbox{\boldmath
  $k$},\mbox{\boldmath $\hat{n}$};\Sigma_{\tau})$ depends only on
$\mbox{\boldmath $\hat{k} \cdot \hat{n}$}$ and $\vert \mbox{\boldmath
  $k$}\vert $ and may therefore be expanded in Legendre polynomials:
\begin{eqnarray}\label{ch4:legexp}
\Delta(\mbox{\boldmath $k$},\mbox{\boldmath
$\hat{n}$};\Sigma_{\tau}) = \sum_{l=0}^{\infty}(-i)^l
(2l+1)\Delta_l(\mbox{\boldmath $k$},\tau) P_l(\mbox{\boldmath
  $\hat{k}\cdot\hat{n}$}) \ .
\end{eqnarray}
The $(-i)^l$ is included by convention to be consistent with most
other authors. \
It is convenient to use the addition theorem at this point to
express the Legendre polynomials in spherical harmonics, giving
\begin{eqnarray} \label{ch4:planewave}
&&\Delta(\mbox{\boldmath $x$},\mbox{\boldmath $\hat{n}$},\tau) = 
4\pi \int d^3k \;e^{i\mbox{\boldmath $k\cdot x$}} 
\sum_{l=0}^{\infty}(-i)^l\;\Delta_l(\mbox{\boldmath $k$},\tau)
\sum_{m=-l}^{l}
Y_{lm}^*(\mbox{\boldmath $\hat{k}$})  
Y_{lm}(\mbox{\boldmath $\hat{n}$}) \ . 
\end{eqnarray}
Employing the orthogonality relation for spherical harmonics $\int
d\Omega \;Y_{lm}^*(\mbox{\boldmath $\hat{n}$})
Y_{l^{\prime}m^{\prime}}(\mbox{\boldmath $\hat{n}$}) =
\delta_{l^{\prime}l}\delta_{m^{\prime}m}$ (where $d\Omega$ is the
integral over solid angle elements centered about direction
$\mbox{\boldmath $\hat{n}$}$) yields 
\begin{eqnarray}
a_{lm}(\mbox{\boldmath $x$},\tau)
&=& \int d\Omega \;\Delta(\mbox{\boldmath $x$},\mbox{\boldmath
$\hat{n}$},\tau) Y_{lm}^*(\mbox{\boldmath $\hat{n}$}) \nonumber \\
&=& (-i)^l\;  4\pi \int d^3 k 
\;e^{i\mbox{\boldmath $k\cdot x$}}
\Delta_l(\mbox{\boldmath $k$},\tau) Y_{lm}^*(\mbox{\boldmath
$\hat{k}$}) \ . \nonumber \\
\end{eqnarray}
The correlation function may now be written as
\begin{eqnarray}\label{ch4:correlfunc}
C_{lm l^{\prime}m^{\prime}}(\mbox{\boldmath $x$},
\mbox{\boldmath $x$}^{\prime})
&=& (-i)^{l-l^{\prime}}\; (4\pi)^2 \int d^3k\;d^3k^{\prime}
e^{i\mbox{\boldmath $k\cdot x$}}e^{-i\mbox{\boldmath $k^{\prime}\cdot
x^{\prime}$}} \nonumber \\
&& \quad\quad\quad\quad\quad\quad
\langle \Delta_l(\mbox{\boldmath $k$},\tau)
\Delta_{l^{\prime}}^*(\mbox{\boldmath $k$}^{\prime},\tau^{\prime})\rangle 
Y_{lm}^*(\mbox{\boldmath $\hat{k}$})
Y_{l^{\prime}m^{\prime}}(\mbox{\boldmath $\hat{k}$}^{\prime}) \ .
\end{eqnarray}

Now since the Boltzmann equation which governs the evolution of the
CMB anisotropy $\Delta_l(\mbox{\boldmath $k$},\tau)$ is independent of
$\mbox{\boldmath $\hat{k}$}$ (in linear theory), 
the $\mbox{\boldmath $\hat{k}$}$
dependence comes entirely from the initial conditions, and we may
write $\Delta_l(\mbox{\boldmath $k$},\tau)$ in terms of the CMB
transfer function $\Delta_l(k,\tau)$ which is defined by 
\citep{1995ApJ...455....7M}:
\begin{eqnarray}
\Delta_l(\mbox{\boldmath $k$},\tau) = \phi_i(\mbox{\boldmath $k$})
\Delta_l(k,\tau) \ ,
\end{eqnarray}
where $\phi_i(\mbox{\boldmath $k$})$ is the initial potential
perturbation and $\Delta_l(k,\tau)$ is real 
By the assumption of translational invariance
$\phi_i(\mbox{\boldmath $k$})$ has a two-point correlation function of
the form
\begin{eqnarray}
\langle \phi_i(\mbox{\boldmath $k$})\phi^*_i(\mbox{\boldmath $k$}^{\prime})\rangle 
= P_{\phi}(k) \;\delta^3(\mbox{\boldmath $k$}-\mbox{\boldmath
  $k$}^{\prime}) \ ,
\end{eqnarray}
where $P_{\phi}(k)$ is the power spectrum of the primordial
(post-inflationary) gravitational potential fluctuations. Then we may
write
\begin{eqnarray}
\langle \Delta_l(\mbox{\boldmath $k$},\tau)
\Delta_{l^{\prime}}^*(\mbox{\boldmath $k$}^{\prime},\tau^{\prime})\rangle 
&=& \Delta_l(k,\tau) \Delta_{l^{\prime}}(k^{\prime},\tau^{\prime}) 
\;P_{\phi}(k) \;\delta^3(\mbox{\boldmath $k$}-\mbox{\boldmath
  $k$}^{\prime}) \ .
\end{eqnarray}

If we evaluate $C_{lm l^{\prime}m^{\prime}}(\mbox{\boldmath $x$},
\mbox{\boldmath $x$}^{\prime})$ for $\mbox{\boldmath
  $x$}=\mbox{\boldmath $x$}^{\prime}$ (and $\tau = \tau^{\prime}$),
the covariance matrix is diagonal and the familiar orthogonality
relation follows
\begin{eqnarray} \label{ch4:Cdiag}
C_{lm l^{\prime}m^{\prime}}(\mbox{\boldmath $x$},
\mbox{\boldmath $x$})
= (4\pi)^2 \delta_{l^{\prime}l}\delta_{m^{\prime}m} \int k^2 dk 
\;\Delta^2_l(k,\tau) P_{\phi}(k) \equiv \delta_{l^{\prime}l}\delta_{m^{\prime}m}
C_l(\tau) \ ,
\end{eqnarray}
where $C_l(\tau)\equiv \langle \vert a_{lm}(\mbox{\boldmath
    $x$},\tau)\vert ^2\rangle$ is the ensemble average of the $l$
harmonic of the CMB power spectrum according to an observer at
conformal time $\tau$.  Thus at any given point all of the $a_{lm}$
are independent random variables.  
Using the addition theorem, we obtain the usual real-space angular
correlation function, at any epoch 
\ba \langle \Delta(\mbox{\boldmath
    $x$},\mbox{\boldmath $\hat{n}$},\tau) \Delta(\mbox{\boldmath
    $x$},\mbox{\boldmath $\hat{n}$}^{\prime},\tau) \rangle  &=&
\frac{1}{4\pi} \sum_{l} (2l+1) C_l(\tau) P_l(\mbox{\boldmath
  $\hat{n}$}\cdot\mbox{\boldmath $\hat{n}$}^{\prime}) \ . 
\ea 
With $\mbox{\boldmath $x$}\ne\mbox{\boldmath $x$}^{\prime}$, $\tau \ne
\tau^{\prime}$ we obtain a more general expression for the correlation
function:
\begin{eqnarray} \label{ch5:moregencorr}
C_{lm l^{\prime}m^{\prime}}(\mbox{\boldmath $x$},\mbox{\boldmath $x$}^{\prime})
&=& (-i)^{l-l^{\prime}}\;  (4\pi)^2 \int d^3k\;
e^{i\mbox{\boldmath $k$}\cdot(\mbox{\boldmath $x$}-\mbox{\boldmath
$x$}^{\prime})} \nonumber \\
&& \quad
\Delta_l(k,\tau) \Delta_{l^{\prime}}(k,\tau^{\prime})
\;P_{\phi}(k) \;Y_{lm}^*(\mbox{\boldmath $\hat{k}$})
\;Y_{l^{\prime}m^{\prime}}(\mbox{\boldmath $\hat{k}$}) \ .
\end{eqnarray}
The symmetries stated in equation (\ref{ch4:symm1}) may be verified from
this expression.  

We may perform the angular part of the $d^3k$ integral in
Eqn.~(\ref{ch5:moregencorr}) by expanding the plane wave piece in spherical Bessel
functions (valid in flat space). 
In a flat FRW space, we are free to manipulate the comoving
3-vectors of events as if we were dealing with position vectors in
Euclidean space (see \emph{e.g.} \cite{1999coph.book.....P}, p.71, Eqn.~(3.19)).
Thus we may define direction vectors $\Delta \mbox{\boldmath $x$} 
\equiv \mbox{\boldmath $x$}-\mbox{\boldmath $x$}^{\prime}$, $\Delta
\mbox{\boldmath $\hat{x}$} \equiv (\mbox{\boldmath
  $x$}-\mbox{\boldmath $x$}^{\prime})/ \vert \mbox{\boldmath
  $x$}-\mbox{\boldmath $x$}^{\prime}\vert$, and expand
\begin{eqnarray}
\exp{\left(i\mbox{\boldmath $k$}\cdot\Delta \mbox{\boldmath
$x$}\right)} 
&=& \sum_{l^{\prime\prime}=0}^{\infty} (2l^{\prime\prime}+1) i^{l^{\prime\prime}} P_{l^{\prime\prime}}(\mbox{\boldmath
$\hat{k}$}\cdot\Delta\mbox{\boldmath $\hat{x}$}) j_{l^{\prime\prime}}(k\vert \Delta \mbox{\boldmath $x$}\vert )
\nonumber \\
&=& 4\pi
\sum_{l^{\prime\prime}=0}^{\infty} i^{l^{\prime\prime}}
j_{l^{\prime\prime}}(k\vert \Delta \mbox{\boldmath $x$}\vert ) 
\sum_{m^{\prime\prime}=-l^{\prime\prime}}^{l^{\prime\prime}}
Y^*_{l^{\prime\prime}m^{\prime\prime}}(\mbox{\boldmath 
$\hat{k}$}) Y_{l^{\prime\prime}m^{\prime\prime}}(\Delta
\mbox{\boldmath $\hat{x}$}) \ ,
\end{eqnarray} 
where we used the addition theorem to separate the $\mbox{\boldmath
  $\hat{k}$}$ and $\Delta \mbox{\boldmath $\hat{x}$}$ dependence.
Then the correlation function becomes
\begin{eqnarray}
C_{lm l^{\prime}m^{\prime}}(\mbox{\boldmath $x$},\mbox{\boldmath $x$}^{\prime})
&=&(4\pi)^3
\int k^2 dk \;
\Delta_l(k,\tau) \Delta_{l^{\prime}}(k,\tau^{\prime}) \;P_{\phi}(k)
\;(-i)^{l-l^{\prime}}\;(-1)^m\;
\nonumber \\
&&
\!\!\!\!\!\!\!\!\!\!\!\!\!\!\!\!\!\!\!\!\!
\!\!\!\!\!\!\!\!\!\!\!\!\!\!\!\!\!
\times\sum_{l^{\prime\prime}m^{\prime\prime}}
i^{l^{\prime\prime}} \;j_{l^{\prime\prime}}(k\vert \Delta \mbox{\boldmath $x$}\vert )
\; Y_{l^{\prime\prime}m^{\prime\prime}}(\Delta\mbox{\boldmath $\hat{x}$})
\int d\Omega_k \;Y_{l,-m}(\mbox{\boldmath $\hat{k}$})
\;Y_{l^{\prime}m^{\prime}}(\mbox{\boldmath $\hat{k}$}) 
Y^*_{l^{\prime\prime}m^{\prime\prime}}(\mbox{\boldmath $\hat{k}$}) 
\ . 
\end{eqnarray}
The angular integral of the product of three spherical harmonics is
expressible in terms of the Wigner $3j$ symbols (see \emph{e.g.}
\cite{Rose}),  
\begin{eqnarray}
&&\int d\Omega_k \;Y_{lm}(\mbox{\boldmath $\hat{k}$})
\;Y_{l^{\prime}m^{\prime}}(\mbox{\boldmath $\hat{k}$}) 
Y^*_{l^{\prime\prime}m^{\prime\prime}}(\mbox{\boldmath $\hat{k}$})
= \sqrt{\frac{(2l+1)(2l^{\prime}+1)(2l^{\prime\prime}+1)}{4\pi}}\;
\biggl(\begin{array}{ccc}
l & l^{\prime} & l^{\prime\prime} \\
0 & 0 & 0
\end{array} \biggr)
\biggl(\begin{array}{ccc}
l & l^{\prime} & l^{\prime\prime} \\
m & m^{\prime} & m^{\prime\prime} 
\end{array} \biggr)
\ .
\end{eqnarray}
The $3j$ symbols are non-zero only if
$m+m^{\prime}=m^{\prime\prime}$, and $l,l^{\prime},l^{\prime\prime}$
satisfy the \emph{triangle condition} that $l^{\prime\prime}$ be equal
to one of $l+l^{\prime}, l+l^{\prime}-1, \cdots, \vert l-l^{\prime}\vert $. The
sum over $l^{\prime\prime}$ therefore reduces to a finite sum. We have finally
\begin{eqnarray}
&&
\!\!\!\!\!\!\!\!\!\!\!\!\!\!\!\!\!\!
C_{lm l^{\prime}m^{\prime}}(\mbox{\boldmath $x$},\mbox{\boldmath $x$}^{\prime})
= \;(-i)^{l-l^{\prime}}\;(-1)^m\; (4\pi)^3
\sqrt{\frac{(2l+1)(2l^{\prime}+1)(2l^{\prime\prime}+1)}{4\pi}}
\nonumber \\
&&\times\sum_{l^{\prime\prime}=\vert l-l^{\prime}\vert
}^{l+l^{\prime}} 
\biggl(\begin{array}{ccc}
l & l^{\prime} & l^{\prime\prime} \\
0 & 0 & 0
\end{array} \biggr)
\biggl(\begin{array}{ccc}
l & l^{\prime} & l^{\prime\prime} \\
-m & m^{\prime} & m^{\prime}-m 
\end{array} \biggr)
\frac{i^{l^{\prime\prime}}}{\sqrt{2l^{\prime\prime}+1}}\;
\;K_{l,l^{\prime},l^{\prime\prime}}(\vert \Delta\mbox{\boldmath $x$}\vert ,\tau,\tau^{\prime})
\;Y^*_{l^{\prime\prime},m^{\prime}-m}(\Delta\mbox{\boldmath
  $\hat{x}$}) \ ,
\end{eqnarray}
where all of the physical information is contained in the kernel
\begin{eqnarray}
&&K_{l,l^{\prime},l^{\prime\prime}}(\vert \Delta\mbox{\boldmath
$x$}\vert ,\tau,\tau^{\prime})  
\equiv \int k^2 dk \; \Delta_l(k,\tau)
\Delta_{l^{\prime}}(k,\tau^{\prime}) \;P_{\phi}(k)\; 
j_{l^{\prime\prime}}(k\vert \Delta \mbox{\boldmath $x$}\vert )
\ .
\end{eqnarray}

\section{Transfer functions \label{ch5:sec_transfer}}

To compute $C_{lml^{\prime}m^{\prime}}$ for a given cosmological
model we need the CMB transfer function $\Delta_l(k,\tau)$.
On the large angular scales accessible via the polarization technique,
the only significant effects responsible for the temperature
anisotropy which need to be included in the transfer function are the
Sachs-Wolfe (SW) and integrated Sachs-Wolfe (ISW) effects 
\citep{1967ApJ...147...73S}. 
The SW effect is the anisotropy due to the gravitational
potential fluctuations on the last scattering surface, and the
associated time dilation effect. The ISW effect arises because, at
late times, as the universe is making the transition from the matter
dominated phase into the vacuum dominated phase, the fluctuations in
the gravitational potential - on scales still in the linear regime -
are still evolving with redshift. As photons fall into and climb out
of this time changing potential they are red-shifted and thus a
temperature anisotropy is generated. 

We first consider the transfer function of the SW effect.
This is computed by ignoring the physics
on scales comparable to the acoustic horizon at the time of
recombination, and retaining only the large scale effects.  In this
limit, the anisotropy is produced solely by the variation in potential
$\phi$ (and the consequent gravitational redshift and time dilation
effects on the photons) and photon density $\delta_{\gamma}$ across
the last scattering shell, ignoring the small scale acoustic waves
which give rise to the acoustic peaks in the angular power spectrum.
Using the line-of-sight integration method \citep{1996ApJ...469..437S},
the SW temperature anisotropy is given in real space by
\begin{eqnarray} \label{ch5:SWlos}
\Delta(\mbox{\boldmath $x$}, \mbox{\boldmath $n$}, \tau) &=&
\int_{0}^{\tau}
d\chi^{\prime}
\;\dot{\zeta}(\tau-\chi^{\prime})\;\left[\frac{1}{4}\delta_{\gamma}(
\chi^{\prime}\mbox{\boldmath$\hat{n}$})  +
\phi(\chi^{\prime}\mbox{\boldmath $\hat{n}$}) \right] \ .
\end{eqnarray} 
Here $\chi^{\prime}$ is the comoving distance measured along the past
light cone of the observer at $(\mbox{\boldmath $x$},\tau)$, in the
direction $\mbox{\boldmath $n$}$.  The visibility function is
defined by $\zeta(\tau)\equiv e^{-\tau_{\rm C}(\tau)}$, with Thomson optical
depth $\tau_{\rm C}(\tau)=\int_{0}^{\tau} d\chi^{\prime}
a(\tau-\chi^{\prime}) n_e(\tau-\chi^{\prime}) \sigma_{\rm T}$ (here and
elsewhere a dot means a derivative with respect to conformal time
$\tau$).  

In the Sachs-Wolfe approximation (valid on scales much
larger than the acoustic horizon), and assuming adiabatic initial
conditions, a perturbative analysis of the equations of motion shows
that $\delta_{\gamma} = -\frac{8}{3} \phi$, and that in Fourier space
the evolution of the potential is given by $\phi(\mbox{\boldmath
  $k$},\tau) = \frac{9}{10} \phi_i(\mbox{\boldmath $k$})$ (see, for
example,  \cite{2002PhRvD..65l3008B,2003Orion.314S..33.}).
The factor of $9/10$ accounts for the evolution of the transfer function between
radiation and matter domination (in the case of adiabatic initial
conditions). Decomposing the plane waves (working here in flat space)
into spherical waves, we obtain 
\begin{eqnarray} 
\Delta_l(k,\tau) &=& \frac{3}{10} \int_{0}^{\tau} d\chi^{\prime}
\;\dot{\zeta}(\tau-\chi^{\prime})\;j_l(k\chi^{\prime}) \ .
\end{eqnarray}
The visibility function $\zeta$ contains the physics of recombination.
It rises rapidly during recombination from $0$ to $1$, with derivative
sharply peaked about the time of recombination, $\tau_r$ \citep{1995ApJ...439..503D}.
The effect of the finite thickness of the last
scattering shell can only influence the radiation field on 
rather small scales, so for times well after recombination, and for
low $l$, we may assume that recombination occurred instantaneously at time $\tau_r$.
In this limit we may take $\zeta$ to
be a delta function centered on $\tau_r$, and the transfer function
reduces to 
\be \label{ch4:swtransfer}
\Delta_l(k,\tau) = \frac{3}{10} j_l[k(\tau-\tau_r)] \ .
\ee
Taking the usual power law form $P_{\phi}(k) = A k^{n-4}$ ($n=1$ gives
the scale-invariant Harrison-Zeldovich spectrum) with an arbitrary
amplitude $A$ (with dimensions of
\newline
$\left(c/H_0\right)^{n-1}$), with the
transfer function in equation (\ref{ch4:swtransfer}) the integral in
equation (\ref{ch4:Cdiag}) may be done analytically, yielding the well
known Sachs-Wolfe expression for the angular power spectrum at low $l$
and $n<3$ (see for example \cite{1999coph.book.....P}): 
\ba 
C_l(\tau) &=& A (4\pi)^2 \left(\frac{3}{10}\right)^2
\int_0^{\infty} k^2 dk \;k^{n-4} j_l^2\left[k(\tau-\tau_r)\right]
\nonumber \\
&=& A(4\pi)^2 \left(\frac{3}{10}\right)^2 \pi 2^{n-4}
\frac{\Gamma(3-n)\Gamma(\frac{2l+n-1}{2})}
{\Gamma^2(\frac{4-n}{2})\Gamma(\frac{2l+5-n}{2})}
\;(\tau-\tau_r)^{1-n} \ .
\ea 
Note that if $n=1$, this expression has no time dependence.
This is a manifestation of the scale invariance property of the
$n=1$ case. 

In the general case including the ISW effect (and assuming a flat
universe), the CMB transfer function is given by a generalization of
Eqn.~(\ref{ch5:SWlos}), the line of sight integral:
\begin{eqnarray}
\Delta(\mbox{\boldmath $x$}, \mbox{\boldmath $n$}, \tau) &=&
\int_{0}^{\tau}
d\chi^{\prime}
\;\dot{\zeta}(\tau-\chi^{\prime})\;\left[\frac{1}{4}\delta_{\gamma}(
\chi^{\prime}\mbox{\boldmath$\hat{n}$})  +
\phi(\chi^{\prime}\mbox{\boldmath $\hat{n}$}) \right] 
\nonumber \\
&& \quad\quad\quad\quad
+ \int_{0}^{\tau} d\chi^{\prime} \;\zeta(\tau-\chi^{\prime})\;
2\dot{\phi}(\chi^{\prime}\mbox{\boldmath $\hat{n}$}) \ .
\end{eqnarray} 
In linear theory, the growth of the amplitude of the potential
perturbations is governed by the growth function $D_+(\tau)$ of the
dark matter perturbation 
The evolution of the potential perturbation in the adiabatic case is
then given by $\phi(\mbox{\boldmath $k$},\tau) = \frac{9}{10}
\phi_i(\mbox{\boldmath $k$}) D_+(\tau)/a(\tau)$.  
\begin{figure*}[tb]
\begin{center}
\includegraphics[width=12cm]{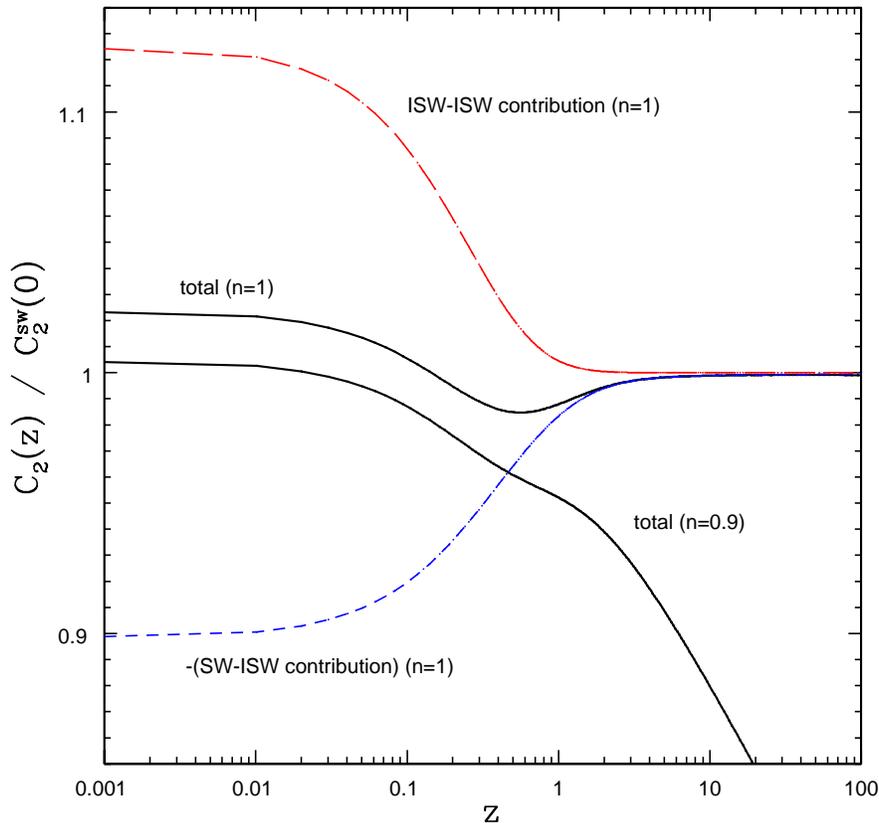}
\end{center}
\caption{A plot of $C_2(z)$, including both the SW
and ISW effects, normalized to the value of the 
SW contribution at $z=0$. The growth function was computed
in a $\Lambda$CDM model with $(\Omega_{\Lambda},\Omega_m)=(0.7,0.3)$.
The upper solid black curve is for the case of power spectral index $n=1$,
and the lower solid black curve for $n=0.9$. The dashed upper and lower curves
curves show the contributions to the $n=1$ case from the ISW-ISW term
and the SW-ISW interference term respectively. These two tend to cancel. 
(Note that the interference term is negative, and its magnitude is plotted here).
\label{ch4:C2}
}
\end{figure*}
In the case of a flat universe with only non-relativistic matter and
vacuum energy, the solution for the growth function, normalized to
$D_+ = a$ at early times, has the simple form
\citep{1977MNRAS.179..351H}:
\begin{eqnarray}
&&D_+(\tau) =
\frac{5}{2} \Omega_m
\frac{\sqrt{\Omega_m+\Omega_{\Lambda}a(\tau)^3}}{a(\tau)^{3/2}}\int_0^{a(\tau)}
X^{3/2}(a^{\prime})\;da^{\prime} \nonumber \\
&&\mbox{where}\;\;\;\;\;\;\;\;\;X(a) =
\frac{a}{\Omega_m+\Omega_{\Lambda}a^3} \ .
\end{eqnarray}
In the instantaneous recombination approximation, this leads to the
following CMB transfer function
\begin{eqnarray}
\Delta_l(k,\tau) = \frac{3}{10} j_l\left[k(\tau-\tau_r)\right] +
\frac{9}{5} \int_{0}^{\tau-\tau_r} d\chi^{\prime} 
\;j_l(k\chi^{\prime})
\; \left.\frac{\partial}{\partial \tau} \frac{D_+}{a}\right\vert
_{\tau-\chi^{\prime}} \ ,
\end{eqnarray}
where the time derivative in the integrand is evaluated at time
$\tau-\chi^{\prime}$.

In the kernel $K_{ll^{\prime}l^{\prime\prime}}$, the product
of two transfer functions appears. Thus there are three terms,
a contribution entirely from the SW effect, an ``interference''
contribution from both the SW and ISW effects, and
a contribution entirely from the ISW effect. Since the ISW
part of the transfer function is usually negative, the
interference term tends to cancel the third term. 
This is illustrated in Fig.~\ref{ch4:C2}, which shows the
redshift dependence of the CMB harmonic $C_2$ in 
a $\Lambda$CDM model with $(\Omega_{\Lambda},\Omega_m)=(0.7,0.3)$,
and various values of the power spectral index $n$.

\section{Scattering of CMB quadrupole \label{ch5:quadscatt}}
 
In this section we consider the generation of polarization by
scattering of the intrinsic CMB quadrupole from electrons
in a galaxy cluster, which we idealize as
a concentrated clump of stationary electrons with a Thomson optical
depth $\tau_{\rm C}$, at a specific angular point on the sky. 

\begin{figure*}[tb]
\begin{center}
\includegraphics[width=13.0cm]{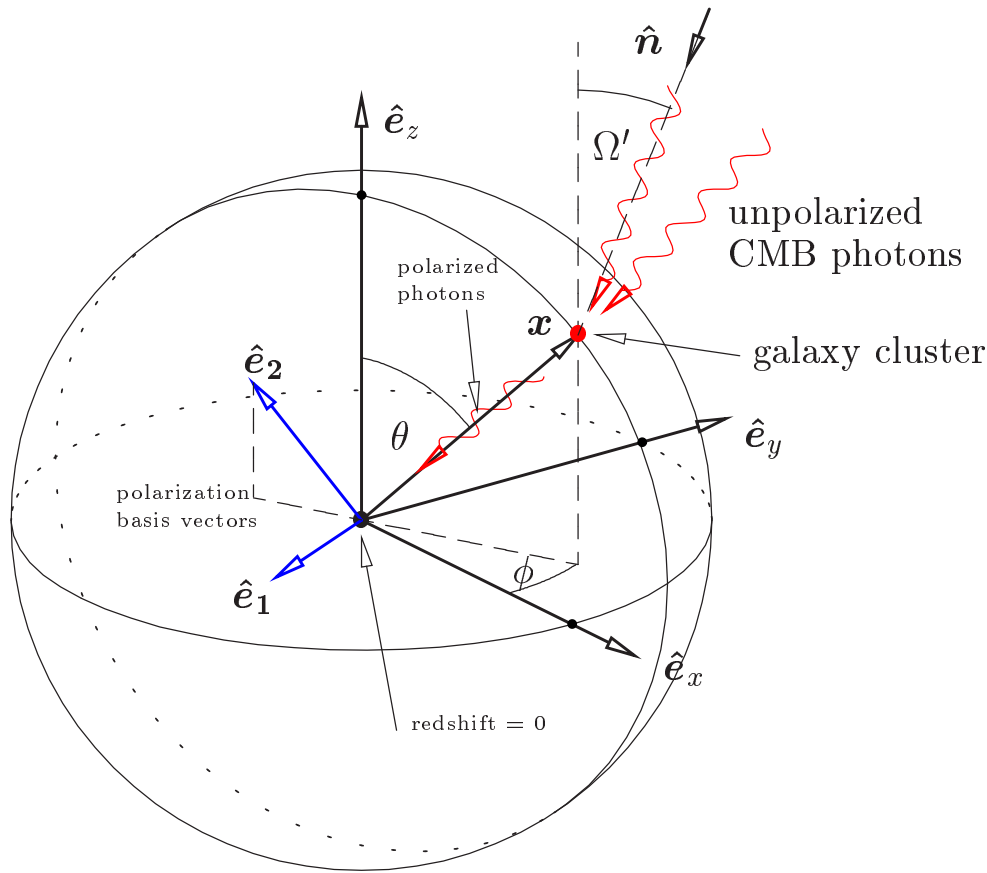} 
\end{center}
\caption{ \label{ch4:galaxy}
Illustrates the coordinate system used to describe the generation
of polarization by Thomson scattering. The CMB incident on a cluster
at comoving position $\mbox{\boldmath $x$}$, which we approximate as
unpolarized, is Thomson scattered and re-radiated by free electrons in the cluster,
producing partially polarized radiation scattered into the line of sight.
At the observer position (redshift $z=0$) the radiation is decomposed
into Stokes parameters with the polarization basis vectors $\hat{\bm{e}}_1,
\hat{\bm{e}}_2$ indicated (defined in equation (\ref{ch5:polbasis}). The CMB radiation incident on the cluster at
$\bm{x}$ is decomposed into spherical harmonics defined with
respect to polar coordinates $\theta'$ and $\phi'$. Only the $l=2$ 
harmonics of the incident radiation field generate polarization.}
\end{figure*}

The Stokes parameters of the radiation scattered into the line
of sight to the cluster are functions of the quadrupole anisotropy in
the local CMB radiation field at the cluster. 
This is characterized by the coefficients $a_{2m}(\bm{x})$ 
of the spherical harmonic expansion of the fractional temperature
anisotropy of the radiation field, which are
functions of the spatial position of the cluster in comoving
coordinates denoted $\bm{x}$ (see \S\ref{ch5:gencorr} for a
description of this coordinate system). 
The direction vector of the line of sight from the observer
at $z=0$ to the cluster is $\mbox{\boldmath $\hat{x}$}$.
The coordinate system used is illustrated in Fig.~\ref{ch4:galaxy}.

In terms of the general set of coefficients $a_{lm}(\bm{x})$, we may
write the brightness temperature of the incident CMB radiation field
at the cluster as a function of the direction vector $\mbox{\boldmath
  $\hat{n}$}$ of the incoming photon \emph{as viewed} from the cluster: 
\be \label{ch4:Iexpand} 
I(\mbox{\boldmath
  $\hat{n}$},\bm{x}) = T_{\rm CMB}(\tau) \sum_{lm} a_{lm}(\mbox{\boldmath $x$})
Y_{lm}(\mbox{\boldmath $\hat{n}$}) \ ,
\ee
where $\tau = \tau_0 - \vert \mbox{\boldmath
  $x$}\vert $ is the conformal time of the scattering events. 
Since the primary anisotropy has a blackbody spectrum, there
is no frequency dependence in $I(\mbox{\boldmath $\hat{n}$},\bm{x})$.

We assume that the incident CMB radiation is unpolarized, 
which  is sufficient to compute the lowest order polarization signal
generated by the quadrupole anisotropy (there are 
relativistic corrections to the effect described here in the case of 
a cluster with a peculiar velocity with respect to the CMB,
as discussed by \cite{2002PhRvD..65j3001C}, which turn out
to be negligible).

The brightness temperature polarization matrix $I_{ij} \sim \langle
E_i E_j \rangle$ of the radiation scattered into the line of sight is
given in the Thomson limit by the following equation: 
\begin{eqnarray} \label{ch5:poltensor}
I_{ij}(\bm{x}) &=& 
\frac{3\tau_{\rm C}}{16\pi} \;
\left(\delta_{ik}-\hat{x}_i\hat{x}_k\right)
\left(\delta_{jl}-\hat{x}_j\hat{x}_l\right)
\int d\Omega^{\prime} \;I(\mbox{\boldmath $\hat{n}$},\bm{x})
\left(\delta_{kl}-\hat{n}_k\hat{n}_l\right) \ ,
\end{eqnarray}
where $d\Omega^{\prime}$ is the solid angle element about the
$\mbox{\boldmath $\hat{n}$}$ direction. A self-contained derivation of
this form of the transfer equation is provided in Appendix \ref{appb}.
The primed solid angle element $d\Omega'$ is associated
with the unprimed direction vector $\hat{\bm{n}}$ since we wish to
reserve $d\Omega$ for the polar angles of the cluster on the sky. 

We now define a polarization basis to define the Stokes parameters
of the radiation incident at the observer from a cluster
in any direction on the sky. We denote the polarization basis vectors
as $\mbox{\boldmath $\hat{e}$}_1, \mbox{\boldmath $\hat{e}$}_2$ 
and leave these unspecified for the moment.  
The Stokes parameters measured in the $\mbox{\boldmath $\hat{e}$}_1,
  \mbox{\boldmath $\hat{e}$}_2$ basis at our position due to scattering in
  the cluster at comoving position $\bm{x}$ are then:
\begin{eqnarray}
I(\mbox{\boldmath $x$})+Q(\mbox{\boldmath $x$}) &=& 
2 I_{ij}(\bm{x}) \;\hat{e}_{1,i} \hat{e}_{1,j} \ , \nonumber \\
I(\mbox{\boldmath $x$})-Q(\mbox{\boldmath $x$}) 
&=&  2 I_{ij}(\bm{x}) \;\hat{e}_{2,i} \hat{e}_{2,j} \ , \nonumber \\
U(\mbox{\boldmath $x$}) &=& 2 I_{ij}(\bm{x}) \;\hat{e}_{1,i} \hat{e}_{2,j} \ .
\end{eqnarray}
Note that we may ignore the Stokes $V$
  parameter  - it remains zero since no circular polarization is
  generated by Thomson scattering.
On substitution of Eqn.~(\ref{ch4:Iexpand}) we find
\begin{eqnarray} \label{ch4:stokes}
Q(\mbox{\boldmath $x$}) &=& 
\frac{3\tau_{\rm C}}{16\pi} T_{\rm CMB}(\tau) \sum_{lm} a_{lm}(\mbox{\boldmath $x$})
\int d\Omega^{\prime} \;Y_{lm}(\mbox{\boldmath $\hat{n}$})\;\left[(\mbox{\boldmath $\hat{e}$}_2\cdot\mbox{\boldmath
$\hat{n}$})^2 -(\mbox{\boldmath $\hat{e}$}_1\cdot\mbox{\boldmath
$\hat{n}$})^2\right] \ ,
\nonumber \\
U(\mbox{\boldmath $x$}) &=& 
-\frac{3\tau_{\rm C}}{8\pi} T_{\rm CMB}(\tau) \sum_{lm} a_{lm}(\mbox{\boldmath $x$})
\int d\Omega^{\prime} \;Y_{lm}(\mbox{\boldmath $\hat{n}$})
(\mbox{\boldmath $\hat{e}$}_1\cdot\mbox{\boldmath $\hat{n}$}) 
(\mbox{\boldmath $\hat{e}$}_2\cdot\mbox{\boldmath $\hat{n}$}) \ .
\end{eqnarray}

To perform the angular integral we need to expand the integrands
in equations (\ref{ch4:stokes}) in spherical harmonics by expressing
$\mbox{\boldmath $\hat{n}$}$ in polar coordinates.
In polar coordinates about the $\mbox{\boldmath $\hat{z}$}$ axis, the
Cartesian components of the direction vectors are taken to be:
\begin{eqnarray}
\mbox{\boldmath  
  $\hat{n}$}&=&
  (\hat{n}_x,\hat{n}_y,\hat{n}_z)=(\sin\theta^{\prime}\cos\phi^{\prime},\sin\theta^{\prime}\sin\phi^{\prime},\cos\theta^{\prime})
  \ , \nonumber \\  
\mbox{\boldmath   
  $\hat{x}$}&=&(\hat{x}_x,\hat{x}_y,\hat{x}_z)=(\sin\theta\cos\phi,\sin\theta\sin\phi,\cos\theta)
  \ .
\end{eqnarray}  
We use the following method.
Any spherical harmonic can be expanded
in terms of the complex quantities $(z_1,z_2,z_3)=(\sin\theta^{\prime}
e^{i\phi^{\prime}},\sin\theta^{\prime}
e^{-i\phi^{\prime}},\cos\theta^{\prime})$
(see for example the discussion of spherical harmonics
in \cite{ByronFuller}).
In terms of these functions we may write $\mbox{\boldmath
  $\hat{n}$}=((z_1+z_2)/2,i(z_2-z_1)/2,z_3)$. 
Note that $z_2 = z_1^*, \;z_3^* = z_3, \;z_1 z_2 = 1-z_3^2$.
It is also convenient to define
\begin{eqnarray}
\mbox{\boldmath $\hat{e}$}_+ &\equiv& \mbox{\boldmath $\hat{e}$}_x +
i\mbox{\boldmath $\hat{e}$}_y \ , \nonumber \\
\mbox{\boldmath $\hat{e}$}_- &\equiv& \mbox{\boldmath $\hat{e}$}_x -
i\mbox{\boldmath $\hat{e}$}_y \ .
\end{eqnarray}
Then for instance we have
\ba \label{ch5:zee}
\mbox{\boldmath $\hat{e}$}_1\cdot\mbox{\boldmath $\hat{n}$} =
\frac{z_1}{2}\;\mbox{\boldmath $\hat{e}$}_1\cdot\mbox{\boldmath $\hat{e}$}_-
+ \frac{z_2}{2}\;\mbox{\boldmath $\hat{e}$}_1\cdot\mbox{\boldmath
  $\hat{e}$}_+ + z_3\;\mbox{\boldmath $\hat{e}$}_1\cdot\mbox{\boldmath
  $\hat{e}$}_z
\ea
The $l=2$ spherical
harmonics may be written as functions quadratic in the
$z$'s as follows:
\begin{eqnarray}
Y_{2,0} &=& \sqrt{\frac{5}{4\pi}} \left(\frac{3}{2} z_3^2 -
\frac{1}{2}\right), \nonumber \\
vY_{2,1} &=& -\sqrt{\frac{15}{8\pi}} \;z_1 z_3, \;\;\;\;\;\;\; Y_{2,-1} = \sqrt{\frac{15}{8\pi}}\; z_2 z_3 \nonumber \\
Y_{2,2} &=& \frac{1}{4}\sqrt{\frac{15}{2\pi}}\; z_1^2, \;\;\;\;\;\;\;
Y_{2,-2} = \frac{1}{4}\sqrt{\frac{15}{2\pi}}\; z_2^2 \ .
\end{eqnarray}

Then we can decompose the integrands into spherical harmonics by
expanding the integrands in equation (\ref{ch4:stokes}) in the $z$'s
using expressions like (\ref{ch5:zee}) and comparing 
with the expressions for $Y_{2m}$ above. 
(Note that in performing this calculation, it is necessary to
use the relation $z_1 z_2 = 1-z_3^2$ to eliminate one of the
coefficients $(z_1,z_2,z_3)$).
We find the following
manifestly real result for the integrands of $Q$ and $U$:
\ba
(\mbox{\boldmath $\hat{e}$}_2\cdot\mbox{\boldmath $\hat{n}$})^2 -(\mbox{\boldmath
$e$}_1\cdot\mbox{\boldmath $\hat{n}$})^2
&=& \sqrt{\frac{8\pi}{5}} \sum_{m=-2}^2 Q_m(\mbox{\boldmath $\hat{x}$})
  Y_{2,m}(\mbox{\boldmath $\hat{n}$}) \ , \nonumber \\
-2(\mbox{\boldmath $\hat{e}$}_1\cdot\mbox{\boldmath $\hat{n}$})(\mbox{\boldmath
$e$}_2\cdot\mbox{\boldmath $\hat{n}$}) 
&=& \sqrt{\frac{8\pi}{5}} \sum_{m=-2}^2 U_m(\mbox{\boldmath $\hat{x}$})
  Y_{2,m}(\mbox{\boldmath $\hat{n}$}) \ .
\ea
The coefficients $Q_m, U_m$ appearing in this expression are 
the following functions of the arbitrary polarization basis vectors chosen
by the observer, which in turn are functions of the cluster
direction on the sky (so $Q_m, \,U_m$ are written as functions
of the cluster direction vector $\hat{\bm{x}}$, which will become
explicit once a polarization basis is chosen):
\begin{eqnarray}
Q_0(\mbox{\boldmath $\hat{x}$}) &=& -\frac{1}{\sqrt{2}}\left[
  (\mbox{\boldmath $\hat{e}$}_1\cdot\mbox{\boldmath $\hat{e}$}_z)^2 -  (\mbox{\boldmath
$e$}_2\cdot\mbox{\boldmath $\hat{e}$}_z)^2 \right] \ , \nonumber \\
Q_1(\mbox{\boldmath $\hat{x}$}) &=& \frac{1}{\sqrt{3}}
  \left[(\mbox{\boldmath $\hat{e}$}_-\cdot\mbox{\boldmath $\hat{e}$}_1)(\mbox{\boldmath $\hat{e}$}_z\cdot\mbox{\boldmath $\hat{e}$}_1)
- (\mbox{\boldmath $\hat{e}$}_-\cdot\mbox{\boldmath $\hat{e}$}_2)(\mbox{\boldmath
  $e$}_z\cdot\mbox{\boldmath $\hat{e}$}_2)\right] \ ,
\nonumber \\
Q_2(\mbox{\boldmath $\hat{x}$}) &=& \frac{1}{2\sqrt{3}}\left[
(\mbox{\boldmath $\hat{e}$}_-\cdot\mbox{\boldmath $\hat{e}$}_2)^2
-(\mbox{\boldmath $\hat{e}$}_-\cdot\mbox{\boldmath $\hat{e}$}_1)^2\right] \ ,
\end{eqnarray}
and
\begin{eqnarray}
U_0(\mbox{\boldmath $\hat{x}$}) &=& 
-\sqrt{2}(\mbox{\boldmath $\hat{e}$}_1\cdot\mbox{\boldmath $\hat{e}$}_z)(\mbox{\boldmath
$e$}_2\cdot\mbox{\boldmath $\hat{e}$}_z) \ , \nonumber \\
U_1(\mbox{\boldmath $\hat{x}$}) &=&
  \frac{1}{\sqrt{3}}\left[(\mbox{\boldmath $\hat{e}$}_-\cdot\mbox{\boldmath
  $e$}_1)(\mbox{\boldmath $\hat{e}$}_z\cdot\mbox{\boldmath $\hat{e}$}_2) +
(\mbox{\boldmath $\hat{e}$}_-\cdot\mbox{\boldmath $\hat{e}$}_2)(\mbox{\boldmath
  $e$}_z\cdot\mbox{\boldmath $\hat{e}$}_1)\right] \ ,
\nonumber \\
U_2(\mbox{\boldmath $\hat{x}$}) &=& 
-\frac{1}{\sqrt{3}}(\mbox{\boldmath $\hat{e}$}_-\cdot\mbox{\boldmath
  $e$}_1)(\mbox{\boldmath $\hat{e}$}_-\cdot\mbox{\boldmath $\hat{e}$}_2) \ .
\end{eqnarray}
Also, we define the quantities with negative $m$ by the relations
\ba
Q_{-m}(\mbox{\boldmath $\hat{x}$}) &\equiv& (-1)^m Q_{m}^{*}(\mbox{\boldmath
  $\hat{x}$})  \ , \nonumber \\
U_{-m}(\mbox{\boldmath $\hat{x}$}) &\equiv& (-1)^m
U_{m}^{*}(\mbox{\boldmath $\hat{x}$}) \ .
\ea

We specialize now to a particular choice of of polarization basis
vectors. A suitable choice is
\begin{eqnarray} \label{ch5:polbasis}
\mbox{\boldmath $\hat{e}$}_1 = \frac{\mbox{\boldmath $\hat{x}$} \times
\mbox{\boldmath $\hat{e}$}_z}{\sqrt{1-\mu^2}} \ , \quad
\mbox{\boldmath $\hat{e}$}_2 = \frac{\mbox{\boldmath $\hat{e}$}_z - \mu
\mbox{\boldmath $\hat{x}$}}{\sqrt{1-\mu^2}} \ ,
\end{eqnarray}
where $\mu = \mbox{\boldmath $\hat{x}$} \cdot \mbox{\boldmath
  $\hat{e}$}_z = \cos\theta$, so that the Stokes parameters $Q, U$ are
  defined with respect to the plane containing the $\mbox{\boldmath
  $\hat{e}$}_z$ axis and the photon direction $-\mbox{\boldmath
  $\hat{x}$}$ (see Fig.~\ref{ch4:galaxy}). 
In this polarization basis the coefficients $Q_m$, $U_m$ are 
\begin{eqnarray} \label{ch5:Qmdef}
Q_0(\mbox{\boldmath $\hat{x}$}) &=& \frac{1}{\sqrt{2}}\sin^2\theta \ , \nonumber \\
Q_1(\mbox{\boldmath $\hat{x}$}) &=& 
\frac{1}{\sqrt{3}}\cos\theta\sin\theta e^{-i\phi} \ , \nonumber \\  
Q_2(\mbox{\boldmath $\hat{x}$}) &=&
\frac{1}{2\sqrt{3}}(1+\cos^2\theta) e^{-2i\phi} \ , \nonumber \\
U_0(\mbox{\boldmath $\hat{x}$}) &=& 0 \ , \nonumber \\
U_1(\mbox{\boldmath $\hat{x}$}) &=& \frac{i}{\sqrt{3}}\sin\theta e^{-i\phi} \ , \nonumber \\ 
U_2(\mbox{\boldmath $\hat{x}$}) &=& \frac{i}{\sqrt{3}}\cos\theta e^{-2i\phi} \ ,
\end{eqnarray}
where $\theta$ is the polar angle between $\bm{x}$ and the $\mbox{\boldmath
  $\hat{e}$}_z$ axis and $\phi$ is the azimuthal angle between the
  projection of $\bm{x}$ on the $(\mbox{\boldmath $\hat{e}$}_x,
  \mbox{\boldmath $\hat{e}$}_y)$ plane
and the $\mbox{\boldmath $\hat{e}$}_x$ axis.

The Stokes parameters may now finally be expressed as a linear
combination of the $a_{2m}(\bm{x})$, with polar axis $\mbox{\boldmath $\hat{e}$}_z$.
Angular integration picks out the five coefficients $a_{2m}$ of the primary anisotropy: 
\ba \label{ch5:QUcluster}
Q(\mbox{\boldmath $x$}) &=& \tau_{\rm C} P_0
\sum_{m=-2}^2 Q_m(\mbox{\boldmath $\hat{x}$}) a_{2m}(\mbox{\boldmath
  $x$}) \ , \nonumber \\
U(\mbox{\boldmath $x$}) &=& \tau_{\rm C} P_0
\sum_{m=-2}^2 U_m(\mbox{\boldmath $\hat{x}$}) a_{2m}(\mbox{\boldmath $x$}) 
\ ,
\ea 
where $P_0 \equiv \frac{3}{4\sqrt{10\pi}} T_{\rm CMB}(\tau)$. Note
that this depends on conformal time, but the fractional distortion in
the Stokes parameters is redshift independent. Also, it turns out that
\be 
Q_m+iU_m = \frac{4}{3}\sqrt{\frac{3\pi}{5}}\; {_2}Y^*_{2m}(\theta,\phi) \ ,
\ee
where ${_2}Y_{lm}$ is the spin-weighted spherical harmonic of spin 2
(see \emph{e.g.} \cite{1997PhRvD..56..596H}).
Thus we may write
\ba
Q(\bm{x}) \pm iU(\bm{x}) = -\frac{6}{20} \sqrt{\frac{2}{3}} \tau_{\rm C} T_{\rm CMB}(\tau)\,
\sum_{m=-2}^2 {_{\pm 2}}Y_{2m}(\mbox{\boldmath $\hat{x}$}) \;a_{2m}(\bm{x}) \ .
\ea

Finally in this section, we quote the following properties of
$Q_m, U_m$, which are needed in \S\ref{ch5:sec_QUstat}
(these are derived using the explicit forms in 
Eqn.~(\ref{ch5:Qmdef})):
\begin{eqnarray} \label{ch5:bidents}
& & \sum_{m=-2}^2 \vert Q_m(\mbox{\boldmath $\hat{x}$}_i)\vert ^2 =
\sum_{m=-2}^2 \vert U_m(\mbox{\boldmath $\hat{x}$}_i)\vert ^2 =
  \frac{2}{3} \ , \nonumber \\
& & \sum_{m=-2}^2 Q_m(\mbox{\boldmath $\hat{x}$}_i)
U^*_m(\mbox{\boldmath $\hat{x}$}_i) = 0 \ , \nonumber \\
& &\sum_{m=-2}^2 (-1)^m Q_m(\mbox{\boldmath $\hat{x}$}_i)
Q^*_m(\mbox{\boldmath $\hat{x}$}_i)
= \frac{2}{3} \left[1-\frac{1}{2}\sin^2 2\theta_i\right] \ , \nonumber \\
& &\sum_{m=-2}^2 (-1)^m U_m(\mbox{\boldmath $\hat{x}$}_i)
U^*_m(\mbox{\boldmath $\hat{x}$}_i) = \frac{2}{3} \cos 2\theta_i \ , 
\nonumber \\
& &\sum_{m=-2}^2 (-1)^m Q_m(\mbox{\boldmath $\hat{x}$}_i)
U^*_m(\mbox{\boldmath $\hat{x}$}_i) = 0 \ .
\end{eqnarray}

\section{Statistics of the cluster polarization signal \label{ch5:sec_QUstat}}

Now we consider the information obtainable from measurements of the
CMB polarization signal (due to scattering of the CMB quadrupole)
from galaxy clusters at various redshifts and lines of sight.  
Assuming that the redshifts of each
cluster can be obtained, this allows mapping, in principle, of a
particular linear combination of $a_{lm}$ over a significant portion
of our past light cone.  Galaxy clusters at similar redshifts and on
lines of sight separated by small angles will produce polarization
signals with Stokes parameters which are strongly correlated.
Widely separated clusters produce uncorrelated signals --- and it is
the combination of these uncorrelated signals that gets around the
cosmic variance bound. 

Using Eqn.~(\ref{ch5:QUcluster}),
the two-point correlation function $\langle Q(\mbox{\boldmath
  $x$})Q(\mbox{\boldmath $x$}^{\prime})\rangle $, of the Stokes
parameters, as defined in the basis Eqn.~(\ref{ch5:polbasis}), 
due to two clusters at general comoving positions $\bm{x},
\bm{x}^{\prime}$ with Thomson optical depths $\tau_{\rm C}, \tau_{\rm C}^{\prime}$ is
given by 
\begin{eqnarray}
  && \!\!\!\!\!\!\!\!\!\!\!\!
\langle Q(\mbox{\boldmath
$x$})Q(\mbox{\boldmath $x$}^{\prime})\rangle 
= \tau_{\rm C} \tau_{\rm C}^{\prime} P_0^2 \sum_{m,m^{\prime}=-2}^2 Q_m(\mbox{\boldmath $\hat{x}$}) 
Q^*_{m^{\prime}}(\mbox{\boldmath $\hat{x}$}^{\prime})\;C_{2m2m^{\prime}}(\mbox{\boldmath
$x$},\mbox{\boldmath $x$}^{\prime}) \ , 
\end{eqnarray}
and similarly for $\langle U(\mbox{\boldmath
    $x$})U(\mbox{\boldmath $x$}^{\prime})\rangle $ and $\langle
    Q(\mbox{\boldmath $x$})U(\mbox{\boldmath $x$}^{\prime})\rangle$. 

\begin{figure*}[tb]
  \centering 
\includegraphics[width=12cm]{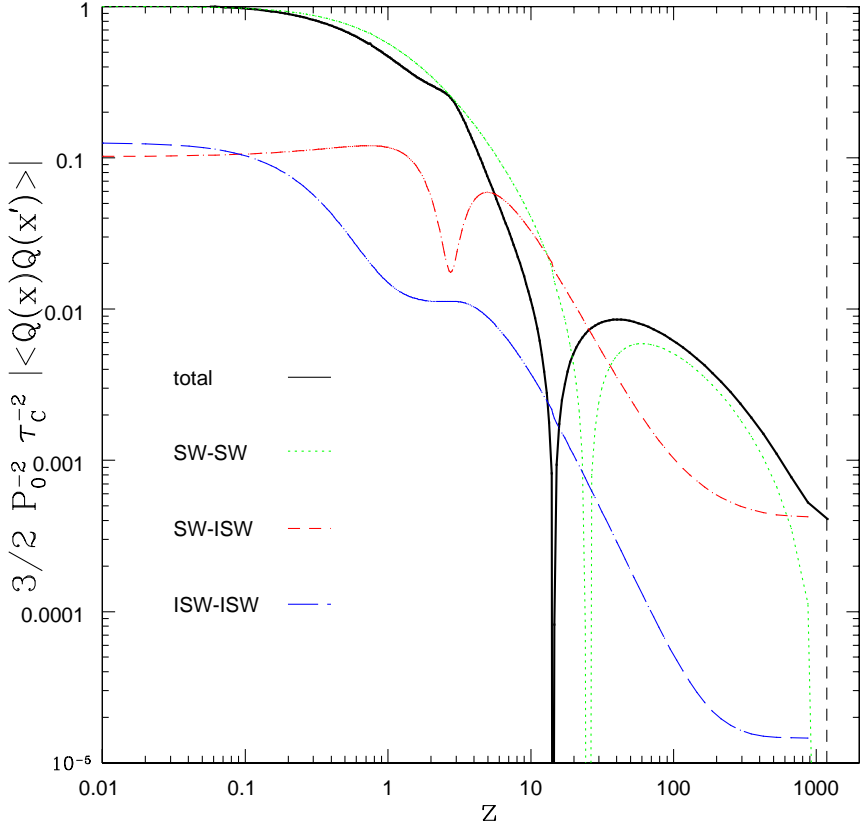}
\caption{
Normalized magnitude of the two-point correlation function of
    the Stokes Q parameter, $\frac{3}{2} P_0^{-2} \tau_{\rm C}^{-2}
    \vert \langle Q(\mbox{\boldmath $x$})Q(\mbox{\boldmath
      $x$}^{\prime})\rangle \vert /C_2(\tau_0)$, where $\bm{x}$ is
    taken to be at redshift $z=0$ and $\bm{x}^{\prime}$ is a point at
    redshift $z$ in the plane orthogonal to $\hat{\bm{z}}$. 
The growth function was computed with cosmological parameters
$\Omega_m=0.35$, $\Omega_{\Lambda}=0.65$, $n=1$. (Note that
the dip in the interference
term, which occurs at $z \approx 3$, is not a zero
crossing).
\label{ch4:qcorrel1}}
\end{figure*}

The two-point correlation function $\langle 
  Q(\mbox{\boldmath $x$}) Q(\mbox{\boldmath $x$}^{\prime})\rangle $,
for points lying on the same line of sight, is shown in 
    Fig.~\ref{ch4:qcorrel1}, which was computed with cosmological
  parameters $\Omega_m=0.35$, $\Omega_{\Lambda}=0.65$, $n=1$.
The solid black curve is the total correlation, and the other curves
  show the contributions from the SW and ISW terms, and their
    interference term. Note that the interference term is negative  - its magnitude is shown here. Only the SW contribution has a genuine zero
crossing. The vertical dashed line indicates the time
  of (instantaneous) recombination. Note that the SW part of
  the correlation passes through zero at redshift $\approx 10$, since
  at redshifts higher than this $\bm{x}^{\prime}$ is in a region of
  the universe separated from the origin by a comoving distance
 greater than $c/(2H_0)$, and thus the $l=2$ correlations die off
  rapidly. 

As $\mbox{\boldmath $x$} \rightarrow \mbox{\boldmath $x$}^{\prime}$,
$\langle  a_{2m}(\mbox{\boldmath $x$}) a^*_{2m^{\prime}}(\mbox{\boldmath
    $x$}^{\prime}) \rangle  \rightarrow C_2(\tau)
\delta_{mm^{\prime}}$, therefore
\begin{eqnarray} \label{ch5:QUvar}
\langle  Q(\mbox{\boldmath $x$})^2 \rangle 
&=& 
\langle  U(\mbox{\boldmath $x$})^2 \rangle 
= \frac{2}{3} \tau_{\rm C}^2 P^2_0 \;C_2(\tau) \ ,
\nonumber \\
\langle  Q(\mbox{\boldmath $x$})U(\mbox{\boldmath $x$}) \rangle 
&=& 0 \ .
\end{eqnarray}
Thus the ensemble average polarization magnitude
due to the scattering of the CMB quadrupole is:
\begin{equation}
\langle \Pi(\bm{x})^2 \rangle \equiv
\frac{\langle  Q(\mbox{\boldmath $x$})^2 \rangle
+ \langle  U(\mbox{\boldmath $x$})^2 \rangle}{T_{\rm CMB}(\tau)^2}
= \frac{4}{3} \tau_{\rm C}^2 \left(\frac{P_0}{T_{\rm CMB}(\tau)}\right)^2
\;C_2(\tau) \ . 
\end{equation}
The quadrupole $Q_2$ is conventionally defined 
by $C_2 = (4\pi/5)Q_2^2$. Thus the root mean square polarization
magnitude is given by (recall $P_0 \equiv \frac{3}{4\sqrt{10\pi}}
T_{\rm CMB}(\tau)$) 
\begin{equation}
\langle P \rangle \equiv \sqrt{\langle \Pi(\bm{x})^2 \rangle} =
\frac{\sqrt{6}}{10} \tau_{\rm C} Q_2 \ ,
\end{equation}
as obtained by \cite{2003PhRvD..67f3505C}.
In a $\Lambda$CDM model, $\langle P \rangle \approx 5\tau_{\rm C}\,\mu$K
(at zero redshift), so the magnitude of this signal is comparable to
that of the other SZ polarization effects. 

Now the relations Eqn.~(\ref{ch5:QUvar})
suggest estimators
$\widehat{C}^Q_2(\tau)$ and $\widehat{C}^U_2(\tau)$ of $C_2(\tau)$, 
given the measured Stokes parameters of $N$ clusters at the same
redshift $z(\tau)$ on lines of sight $\hat{\bm{x}}_i$ with optical
depths $\tau_{C,i}$ $(i=1,\cdots\!,N)$, which beat the cosmic variance
limit: 
\begin{eqnarray} \label{ch5:c2est}
  \widehat{C}^Q_2(\tau) &\equiv& \frac{3}{2} P^{-2}_0 \sum_i W_i \;
  Q(\bm{x}_i)^2 \ , \nonumber \\
  \widehat{C}^U_2(\tau) &\equiv& \frac{3}{2} P^{-2}_0 \sum_i W_i \;
  U(\bm{x}_i)^2 \ .
\end{eqnarray}
Note that these are non-optimal, coordinate-dependent estimators.
In future work it would be interesting to construct
optimal estimators of the power at a given scale, given the cluster
polarization data, but we do not attempt that here. 
For the mean of these estimators to equal $C_2(\tau)$, 
the weights $W_i$ must be chosen to satisfy
\be
\sum_i W_i \; \tau^2_{C,i} = 1 \ .
\ee
We will consider the simple choice 
\be \label{ch4:simpleest}
W_i = \frac{\tau_{C,i}^{n-2}}{\sum_j \tau_{C,j}^{n}} \ .
\ee
Note that $n=0$ would be a bad choice since it gives
more weight to clusters producing weaker signals,
reducing the signal to noise. A
better choice is the uniform weighting $n=2$.

The cosmic variance limit on these estimators is determined by the variances (with $X,X^{\prime}$ indicating either $Q$ or $U$)
\ba
\mbox{Var} \;\widehat{C}_2^{X}(\tau)
&=& \frac{9}{4} P_0^{-4} \sum_{ij} W_i\; W_j\;
\langle X(\bm{x}_i)^2 X(\bm{x}_j)^2\rangle \ .
\ea 
The sum over $i,j$ may be broken into a 
contribution from clusters at the same location, 
$\mbox{Var}_1(X)$, and a 
contribution from clusters at separate locations, 
$\mbox{Var}_2(X,X)$:
\begin{eqnarray} \label{ch4:defvar}
&&\mbox{Var} \;\widehat{C}^X_2(\tau) 
= \mbox{Var}_1(X) + \mbox{Var}_2(X,X) \ ,
\end{eqnarray}
where
\begin{eqnarray} \label{ch4:estvar}
&&\mbox{Var}_1(Q)
= P_0^{-4} C_2(\tau)^2  \sum_i W_i^2 \tau_{C,i}^4 \left[ 1 + (1-\frac{1}{2}\sin^2 2\theta_i)^2 \right]  \ ,
\nonumber \\ 
&&\mbox{Var}_1(U)
= P_0^{-4} C_2(\tau)^2  \sum_i W_i^2 \tau_{C,i}^4 \left[ 1 + \cos^2 2\theta_i \right] \ ,
\end{eqnarray}
and
\begin{eqnarray} \label{ch4:deltavar}
\mbox{Var}_2(X,X^{\prime}) = \frac{9}{2}
\sum_{i,\;j>i} W_i\;W_j\;\tau^2_{C,i}\tau^2_{C,j} \left[
\;\left|\sum_{m_1 m_2} 
X_{m_1}(\mbox{\boldmath $\hat{x}$}_i) 
X^{\prime *}_{m_2}(\mbox{\boldmath   $\hat{x}$}_j) \;C_{2m_1
  2m_2}(\mbox{\boldmath $x$}_i,\mbox{\boldmath $x$}_j) \right|^2  
\right.
&&
\nonumber \\ 
\left. \quad\quad\quad\quad\quad\quad\quad\quad
+ \left|\sum_{m_1 m_2} (-1)^{m_2}
X_{m_1}(\mbox{\boldmath $\hat{x}$}_i) 
X^{\prime *}_{m_2}(\mbox{\boldmath $\hat{x}$}_j)  \;C_{2m_1
  2,m_2}(\mbox{\boldmath $x$}_i, \mbox{\boldmath $x$}_j)\right|^2 
\right]
\ .
\end{eqnarray}
Here we used the 4-point correlation function from
Eqn.~(\ref{ch5:4point}), the relations in equations
(\ref{ch5:bidents}), and the fact that
$C_{2m2m^{\prime}}(\mbox{\boldmath $x$}_i,\mbox{\boldmath $x$}_i) = 
\delta_{mm^{\prime}} C_2(\tau)$.
Note that the variances are functions of the angular positions
of the clusters on the sky, which is due to the specific choice of
polarization basis for the Stokes parameters. 

\begin{figure*}[t]
\begin{center}
  \includegraphics[width=8.5cm]{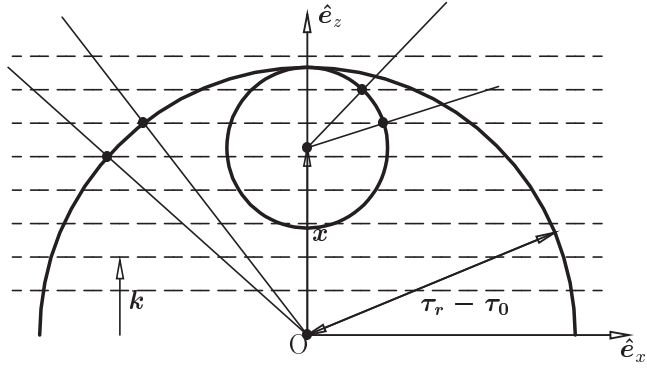}
\caption{Illustrates that the angular scale of a given comoving $k$-mode 
  subtended on the CMB sky of an observer at high redshift is greater
  than the angular scale of the same $k$-mode on the CMB sky of an
  observer at low redshift.
\label{ch5:angscale}}
\end{center}
\end{figure*}

If all of the correlations $C_{2m2,m^{\prime}}(\mbox{\boldmath
  $x$}_i,\mbox{\boldmath $x$}_j)$ with $i\ne j$ vanish, then only the
first term on the right hand side of equation (\ref{ch4:defvar})
remains. If however the cluster positions $\bm{x}_i$ are close enough
that the these correlations approach $C_2(\tau)$, then the $O(N^2)$
terms in the $\mbox{Var}_2(X,X^{\prime})$ terms combine to swamp the first terms.
Thus in order to beat cosmic variance by a factor of $O(N^{-1/2})$ we
need $N$ sets of clusters which are mutually uncorrelated
(as pointed out in a qualitative discussion by
\cite{Kamionkowski:1997na}).
The number of uncorrelated regions available increases as the redshift
increases, since the comoving region surrounding each cluster outside
of which the polarization is approximately uncorrelated 
with that produced by the cluster is smaller at higher redshift. 
This is because smaller comoving scales contribute to the $l=2$
harmonic of the CMB on the sky at higher redshift,
and similarly the same comoving scale
maps into different angular scales, as illustrated in
Fig. \ref{ch5:angscale}. 
$C_2(\tau<\tau_0)$ depends on fluctuations of smaller scale
than $C_2(\tau_0)$.

At very low redshift, any cluster
will be correlated with any other, and we get back the usual cosmic
variance constraint.
In other words, we can beat the cosmic variance on $C_2(\tau<\tau_0)$,
but not on $C_2(\tau_0)$, today's quadrupole.

To demonstrate the reduction in cosmic variance, 
we first combine the $Q$ and $U$ measurements
to obtain an improved estimator of $C_2(\tau)$.
Taking a linear combination of $\widehat{C}^Q_2, \widehat{C}^U_2$
yields an improved estimator,
\begin{eqnarray}
  \widehat{C}^P_2(\tau) &\equiv& \alpha \widehat{C}^Q_2(\tau)
+ (1-\alpha) \widehat{C}^U_2(\tau) \ ,
\end{eqnarray}
with $0 \le \alpha \le 1$. This has variance
\begin{eqnarray} \label{ch4:optvar}
\mbox{Var} \;\widehat{C}^P_2(\tau) &=& \alpha^2 \mbox{Var}
  \;\widehat{C}^Q_2(\tau) + (1-\alpha)^2 \mbox{Var}
  \;\widehat{C}^U_2(\tau) \nonumber \\
&& \quad\quad
+ 2\alpha(1-\alpha) \;\mbox{Cov}(\widehat{C}^Q_2,\widehat{C}^U_2) \ .
\end{eqnarray}
It is easy to see that the covariance
$\mbox{Cov}(\widehat{C}^Q_2,\widehat{C}^U_2)$ is zero
because of the relations in equations (\ref{ch5:bidents}),
which imply $\mbox{Var}_2(Q,U)=0$.
Thus the optimum value of $\alpha$ is trivially $\alpha=1/2$.

\begin{figure*}[tb] 
\mbox{\subfigure[\hspace{0.1cm}$\mbox{Var}\;C_2(z)$]{\includegraphics[width=8cm,height=7.5cm]{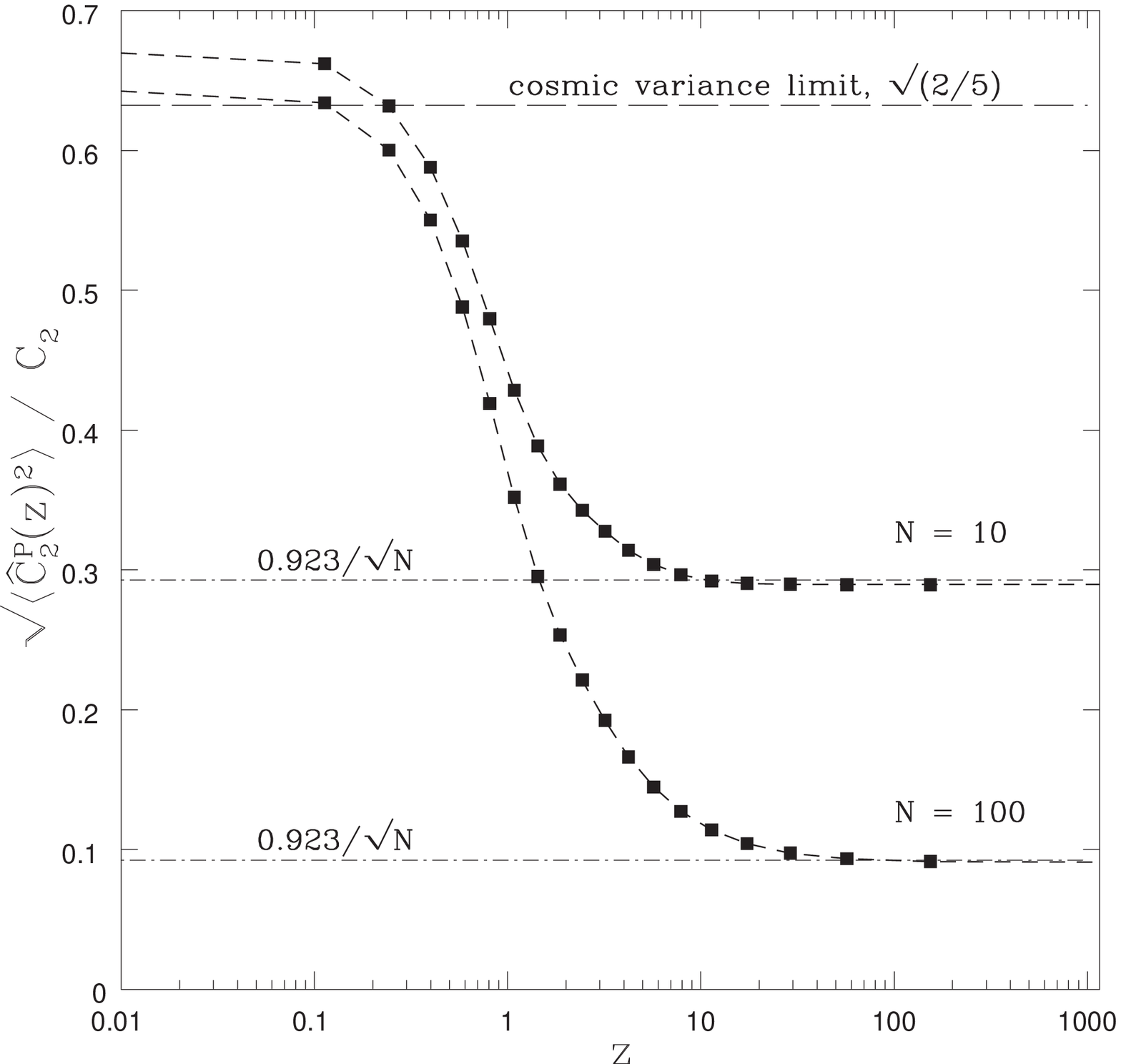}}
\quad 
\subfigure[\hspace{0.1cm}Ratio of $\mbox{Var}\;C_2(z)$ to
  $\mbox{Var}\;C_{l_{\rm{eff}}}(z)$]{\includegraphics[width=8cm,height=7.5cm]{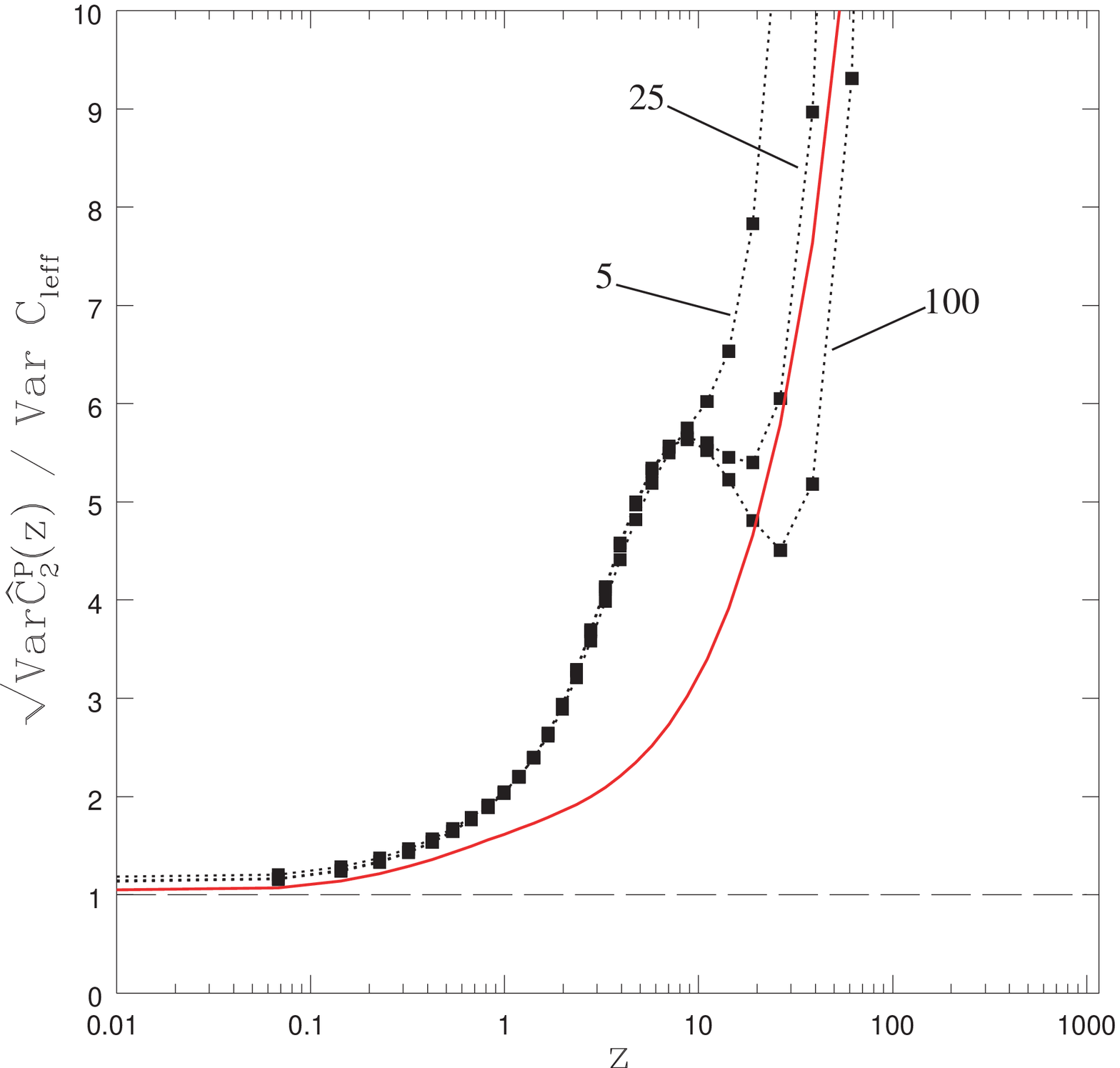}}}\caption{
Reduction in cosmic variance with clusters distributed on various
redshift shells in random directions. 
\label{ch5:cosmicvar2}    
}
\end{figure*}

With these expressions we can compute the variance
of our estimator for $C_2$ given a set of cluster
positions and optical depths $\{\mbox{\boldmath
  $x$}_i\},\;\{\tau_{\rm{C},i}\}$. For simplicity, we will compute
here only the variances for sets of clusters which lie on a given
redshift shell, distributed in random directions.
In the left panel of Fig.~\ref{ch5:cosmicvar2} we show the variance of
the estimator for $C_2$ obtained from hypothetical cluster
polarization measurements from sets of $10$ and
$100$ clusters, \emph{all on the same shell of redshift $z$},
distributed in random directions on the sky.
(We assume the cosmological parameters $\Omega_m=0.35,
\;\Omega_{\Lambda}=0.65$, power spectral index $n=1$).
In the left panel, we show the variance of the estimator
for $C_2$, as a fraction of $C_2$ at $z=0$,
from separate sets of clusters confined to redshift shells
as a function of the redshift of the shell.

For simplicity we have assumed that all clusters have the same optical
depth $\tau_{\rm C}$, so the weights given in equation
(\ref{ch4:simpleest}) are all equal, $W_i = \tau_{\rm C}^{-2}/N$.
The filled squares indicate the square root of
$\mbox{Var}\;\widehat{C}^P_2$, the variance of the average estimator
for $C_2$, $(\widehat{C}^Q_2 + \widehat{C}^U_2)/2$, for $10$ and $100$
clusters as indicated. This is 
expressed as a fraction of the Sachs-Wolfe contribution to
$C_2$, which is independent of redshift for $n=1$. 
The horizontal dashed line at $\sqrt{2/5}\approx 0.63$ indicates
the cosmic variance limit on a single CMB sky given in
Eqn.~(\ref{ch5:cvar}).
(Note that the ISW contribution is included in these calculations,
but omitting it leads to a difference of less than a few percent
in the curves in Fig.~\ref{ch5:cosmicvar2}).
As $z \rightarrow 0$, the estimator variance slightly
exceeds the cosmic variance limit given a single CMB sky ---
clusters with overlapping correlation spheres as $z\rightarrow 0$
are no more useful than a direct measurement of the quadrupole on our
sky by \emph{e.g.} WMAP. In fact our estimator of $C_2$ is worse than a direct measure
 --- but an optimal estimator could
be constructed which would yield all of the $a_{2m}$ at $z\approx 0$
from cluster measurements at various points on the sky.
As $z \rightarrow \infty$, the estimator variance 
goes as $\sim 1/\sqrt{N}$
(as $z \rightarrow \infty$, the variance 
approaches $0.923/\sqrt{N}$. This may be derived by averaging the
estimator variances in Eqn.~(\ref{ch4:estvar}) 
over angles in the $N\rightarrow\infty$ limit).

Thus if we have enough clusters, we can
measure $C_2(z)$ at increasingly high accuracy as the redshift
shell increases. 
However the fact that $C_2(z>0)$ can be measured more accurately
than $C_2(z=0)$ does not necessarily mean that we can reconstruct the
primordial power more accurately than we could using only say the WMAP
multipoles, unless the variance in our estimator
is less than the variance in the CMB multipole which probes
the power at the scale corresponding to the quadrupole at redshift $z$.

In the right panel of Fig.~\ref{ch5:cosmicvar2} we show the variance
of our estimator of $C_2$, except divided through by the cosmic
variance in the CMB multipole $C_{l_{\rm{eff}}}$, where $l_{\rm{eff}}$
is the $l$-scale on our sky corresponding to the $l=2$ scale on the
redshift shell at $z>0$.  
To be explicit, we are plotting
\begin{equation} \label{keyplot}
\frac{\sqrt{\mbox{Var}\;\widehat{C}_2}}{\sqrt{2/(2 l_{\rm{eff}}+1)}
  \;C_{l_{\rm{eff}}}} \ .
\end{equation}
This is the key plot, since if this ratio is less than
unity, we have shown that this method can improve on the
cosmic variance limits inherent in the existing CMB data,
as discussed in the introduction.
A reasonable approximation for $l_{\rm{eff}}$ is to take the real $l$
value corresponding to the comoving radius of the last scattering
sphere at the redshift $z$, namely
$l_{\rm{eff}}=2/(1-\chi/(\tau_0-\tau_r))$ where $\chi(z)$ is the
comoving distance of the shell.
Note that there is no one length scale or ($z=0$) harmonic which
corresponds to the ($z>0$) quadrupole, since a single angular harmonic
mode has power spread over a range of scales.   
In Appendix \ref{testleff} we check the approximation for
$l_{\rm{eff}}$ given above by considering the $k$-space window
function of the quadrupole, and find that a slight modification to
the formula for $l_{\rm{eff}}$ given above gives a better
approximation. We use this modified formula in our subsequent
calculations. We opt to compute $l_{\rm{eff}}$ as a real number and 
interpolate between the $C_l$'s at the integer values of
$l$ bracketing $l_{\rm{eff}}$, to obtain the corresponding cosmic
variance. This will give a reasonably good approximation to the
accuracy achievable in power spectrum estimation.

A rough estimate of the maximum number $N_{\mbox{max}}$ of clusters
with uncorrelated signals available on a given redshift shell is given
by dividing the area of the shell by the area of a circle
with radius equal to that of the last scattering sphere at that
redshift: 
\begin{equation}
N_{\mbox{max}} = 1 + 4\chi^2/(\tau_0-\tau_r-\chi)^2 \ ,
\end{equation}
where $\chi(z)$ is the comoving distance to the shell.
The first term is added to allow for the fact that at least one
independent cluster signal is available at low redshift.
An approximation to the minimum estimator variance achievable on a
given redshift shell is then given by:
\begin{equation}
\frac{\sqrt{2/5}\;C_2(z=0)}{\sqrt{N_{\mbox{max}}}} \ .
\end{equation}
An approximate lower bound on the ratio of the square root of
the estimator variance and the cosmic variance in $C_{l_{\rm{eff}}}$
is thus given by:
\begin{equation}
\frac{C_2(z=0)}{C_{l_{\rm{eff}}}}\sqrt{\frac{2l_{\rm{eff}}+1}{N_{\mbox{max}}}} \ .
\end{equation}
This is shown by the solid curve in the right panel
of Fig.~\ref{ch5:cosmicvar2}.
Since this curve exceeds $1$ for all values of $z$, this suggests that
the cosmic variance is not reducible below the WMAP levels with the
cluster technique even without computing the detailed correlations. 

In the right panel of Fig.~\ref{ch5:cosmicvar2}, the black squares
joined by dashed lines show the ratio (\ref{keyplot}) computed with
the full expression for the estimator variance, plotted for sets of
$5$, $25$ and $100$ clusters as indicated. 
Clearly there is no reduction in variance below the WMAP
cosmic variance at any redshift with up to $100$ clusters,
as expected from our rough analysis.
The three curves are barely distinguishable up to $z\approx 10$,
reflecting the fact that at low redshift increasing the number
of clusters does not enhance the reduction in cosmic variance
since the polarization signals are correlated. 
Increasing the number of clusters further will not lead
to significant improvement, at least with signals from $z<10$.
For $z>10$, the $N=25$ and $N=100$ cases begin to show a drop
in the ratio, as the redshift shell expands (and last scattering
spheres around each cluster shrink) to the point at which 
the signals from each cluster are uncorrelated.
As $z$ increases eventually the ratio in the $N=25$ and $N=100$ cases
begins to grow again, since $l_{\rm{eff}}$ is increasing as the last
scattering spheres shrink, and so the cosmic variance in
$C_{l_{\rm{eff}}}$ is decreasing.  

Unfortunately it seems that the cluster technique,
at least using the estimator we propose, is not
competitive with the CMB data at $z=0$ for estimation of the power at
a given scale. 
In future work it would be interesting
to attempt to construct a better estimator
--- for example combining signals from sets of clusters
distributed at arbitrary redshift rather than on redshift shells
would presumably lead to some improvement.
However we emphasize that this technique is still a useful probe
of the time evolution of $C_2$, and possibly the ISW effect.

\section{Discussion \label{ch5:sec_discuss}}

We have developed a statistical theory of the part of the polarization
signal in the CMB in the direction of galaxy clusters produced
by scattering of the CMB temperature quadrupole.
We have shown explicitly that it is possible
to use the indirect information about the last scattering
surfaces of distant observers contained in these polarization
measurements to constrain the $l=2$ angular power spectrum harmonic of
the CMB, $C_2$,  as a function of redshift with greater statistical
accuracy.  
We also showed however that it not possible to use the cluster
polarization measurements to probe the power on a given scale
with higher accuracy than the limits imposed by cosmic variance
on the single sky CMB data. Thus power spectrum estimation
cannot be improved using this method.

But we believe the cluster method is still of considerable value though,
since it serves as a probe of the physical mechanism which might
have suppressed the quadrupole. It has been noted that the
quadrupole seems to be lower than might be expected due to a
statistical fluctuation alone \citep{deOliveira-Costa:2003pu}.
The quadrupole may be anomalously low, that is lower than the standard
models predict, for various reasons: a cutoff at large scales in the
fluctuation power spectrum, different-than-expected behavior of the
transfer function at large scales, or effects associated with the
large-scale topology or geometry of the universe.
There could conceivably be other explanations, but these are the
simplest. Signatures of these effects could be present in the time
evolution of $C_2$.
For example if the quadrupole is low because of some
topological suppression, we should see a rise in $C_2$ with
redshift as the scale probed falls below the local quadrupole scale
(but if the standard models are correct, there would be no such rise, at
least in the $n=1$ case). 

One might be worried that since the accuracy of the cluster
measurement of $C_2$ is not higher than the accuracy of the
measurement of the corresponding WMAP harmonic, 
as shown is \S\ref{ch5:sec_QUstat}, the cluster method may
not provide any more information that that already contained in the
WMAP data.
However the cluster technique yields
information which is not obtainable with WMAP (about perturbations on last
scattering surfaces different to our own)
and is thus complementary.
In future work it would be desirable to perform a full analysis of the
feasibility of using the cluster method to detect the effects of
topology (or other quadrupole suppression mechanisms) in the time
evolution of $C_2$, and a comparison with what we can already learn
from the WMAP data. 
 
We also showed that in the standard $\Lambda$CDM model
the ISW effect produces a small ($\approx 2$\%) bump in CMB harmonic
$C_2(\tau)$, which is swamped by the high cosmic variance at low
redshift even using large numbers of clusters. 
However the ISW effect produces
a significant feature in the two-point correlation function of the
Stokes parameters which might be detectable. Detection of the ISW
effect would provide additional information about the
acceleration of the universe and the dark energy.
We also note that this method is a rather sensitive probe of
deviations from the scale-invariant power spectrum, because if $n\ne
1$ then the Sachs-Wolfe contribution to $C_2(\tau)$ either grows or decays
rapidly with conformal time.

The procedure we have outlined is something of an
idealization. We assumed that the polarization signal induced by the
quadrupole is obtainable from many clusters at the same redshift, and
we ignored noise and contamination of the signal.  
In practice there are contaminating polarization signals from the
kinetic and thermal Sunyaev-Zeldovich effects \citep{Diego:2003dp},
and distortions in the polarization field due to lensing
\citep{2000ApJ...538...57S}. 
Clearly separating the quadrupole signal from the other contaminants
would be a major experimental challenge. 
However the signal to noise may be increased
to some extent by combining signals from clusters located at similar
directions and redshifts, since the signal from sufficiently nearby
clusters is strongly correlated.
It remains to be seen if this technique will be a practically useful
cosmological probe.

\acknowledgements
\noindent
I thank Pedro Ferreira, Joseph Silk, 
and Constantinos Skordis for helpful discussions.

\appendix

\section{Transfer equation for polarization matrix \label{appb}} 

We found it convenient to write the transfer equation for
generation of polarization by Thomson scattering in matrix form.
This approach is similar to the ``density matrix'' formalism
for polarized radiation transfer \citep{1994PhDT........24K,2000PhRvD..62d3004C}.

The transfer equation is usually written in terms of the Stokes
parameters, which are time averages
of quadratic products of the complex amplitudes of the electric field
components $E_i$ of the electromagnetic wave. 
The polarization state and intensity of a beam of light propagating in
the $z$--direction is characterized completely by the $2\times 2$
Hermitian matrix $\langle E_i E^*_j \rangle$, with
$(i,j)\in\{x,y\}$ (the brackets denote a time average).
An obvious generalization is to allow $(i,j)$ to
become Cartesian tensor indices and to run over all of $\{x,y,z\}$. We
obtain a $3\times 3$ polarization matrix associated with photon direction $\bm{n}$: 
\begin{equation}
Q_{ij}(\bm{n}) = \left< E_i E_j^*\right> \ . \quad\quad (i,j) \in \{x,y,z\}
\end{equation}
With this matrix we are no longer restricted to considering a beam
propagating along a coordinate axis. 
For a given photon direction $\bm{n}$, the electric fields are
transverse to $\bm{n}$, implying
\begin{equation}
n^i Q_{ij}(\bm{n}) = 0 \ .
\end{equation}
Consider now a superposition of beams of various directions
$\bm{n}$ and momenta $\bm{p}=p \bm{n}$, where $p$ is the frequency
(we set $c=h=1$ for convenience). In this case, $Q_{ij}$ cannot be
considered a function of the photon momenta, but an intensity matrix
can be defined associated with each photon direction and frequency.
Recall from the definition of specific intensity that the energy
density is given by $Q^i_i=\int I(\bm{p})\,dp 
\,d\Omega$, where $I(\bm{p})$ is the specific intensity and $d\Omega$
is the solid angle element associated with the photon direction $\bm{n}$.
Similarly we can express each element of the matrix $Q_{ij}$ in
terms of specific intensity matrices $I_{ij}$:
\begin{equation} \label{ch1:def_fmunu}
  Q_{ij} = \int dp\;d\Omega\; I_{ij}(\bm{p}) \ .
\end{equation}

Now the Stokes parameters are defined with respect to a particular choice
of ``polarization basis''. This is a pair of mutually orthogonal unit
vectors $\bm{\epsilon}^{(1)}, \bm{\epsilon}^{(2)}$, both orthogonal to the beam
direction.
The Stokes parameters are given in terms of $I_{ij}$ and the
polarization basis vectors as:
\begin{eqnarray} \label{ch1:stokesbasis}
\frac{I+Q}{2} &=& \epsilon^{(1)}_i \epsilon^{(1)}_j I_{ij} \ , \nonumber \\
\frac{I-Q}{2} &=& \epsilon^{(2)}_i \epsilon^{(2)}_j I_{ij} \ , \nonumber \\
\frac{U+iV}{2} &=& \epsilon^{(1)}_i \epsilon^{(2)}_j I_{ij} \ , \nonumber \\
\frac{U-iV}{2} &=& \epsilon^{(2)}_i \epsilon^{(1)}_j I_{ij} \ .
\end{eqnarray}
In the case of
a beam propagating in the $z$-direction for example, we have, choosing
polarization basis vectors $\bm{\epsilon}^{(1)}=\bm{x}, \;\bm{\epsilon}^{(2)}=\bm{y}$,
\begin{equation} \label{Qofzbeam}
I_{ij} = \frac{1}{2} \left( \begin{array}{ccc}
I+Q & U+iV & 0 \\
U-iV & I-Q & 0 \\
0 & 0 & 0 
\end{array} \right) \ . 
\end{equation}
The matrix is non-zero only in the two dimensional subspace spanned by
$\bm{\epsilon}^{(1)}, \bm{\epsilon}^{(2)}$.

We need to construct the matrix of an unpolarized beam propagating in
a general direction $\bm{n}$. The only quantities available to form
this matrix are the intensity $I$, the components of the direction
vector $\bm{n}$, and the Kronecker delta $\delta_{ij}$. 
The matrix must therefore be of the form:
\begin{equation} \label{ch1:unpoltensorform}
I_{ij}(\bm{n}) = A \delta_{ij} + B n_i n_j \ .
\end{equation}
The matrix of an unpolarized beam propagating in the $z$--direction
is obviously
\begin{equation}
I_{ij} = \frac{I}{2} \left(\begin{array}{ccc}
1 & 0 & 0 \\
0 & 1 & 0 \\
0 & 0 & 0 \end{array}\right) .
\end{equation}
Comparing this with the form of Eqn.~(\ref{ch1:unpoltensorform}) 
for the special case $n_i = \delta_{iz}$, we see that $A=-B=I/2$.
Thus the matrix of an unpolarized beam in a general direction $\bm{n}$
is
\begin{equation}\label{appb:unpol}
I_{ij}(\bm{n}) = \frac{I}{2} (\delta_{ij}-n_i n_j) \ .
\end{equation}
The polarization magnitude of the beam described by a general matrix
$I_{ij}$ is given by
\begin{equation}
\Pi^2 = 2 \mbox{Tr}[\bm{I}^2]/\mbox{Tr}[\bm{I}]-1 \ ,
\end{equation}
which may be readily checked with the matrix (\ref{Qofzbeam}).

We now derive the equation for the time evolution of the polarization
matrix due to Thomson scattering from a distribution of stationary
electrons. For a completely linearly polarized beam,
$Q^{ij}\epsilon^\ast_i \epsilon_j$ is the time-average energy
density for electromagnetic radiation of polarization $\bm\epsilon$, where
$\bm{\epsilon}\cdot\bm{\epsilon}=1$. Consider a completely polarized
beam with polarization vector $\bm{\epsilon}$ and momentum
$\bm{p}=p\bm{n}$ incident upon an electron at rest 
The polarization matrix of the incident beam is
$Q_{ij}=Q \epsilon_i \epsilon_j^{\ast}$ where $cQ$ is the
incident flux (we choose units such 
that $c=1$). Normalization of the polarization vector implies
$Q=Q_{ij}\epsilon^\ast_{i}\epsilon_{j}$. In the
Thomson limit, in which the electron recoil is negligible, 
the differential cross section for Thomson scattering of
a beam into final momentum $\bm
p'=p'\bm{n}'$ and polarization $\bm\epsilon'$ is \citep{Jackson}
\begin{eqnarray}\label{ch3:thomxsec}
  \frac{d\sigma}{d\Omega'}=\frac{3\sigma_{\rm T}}{8\pi}
\left\vert\bm\epsilon^{\ast}\cdot\bm\epsilon'\right\vert^2 \ .
\end{eqnarray}
where $d\Omega'$ is the solid angle element associated with
the scattered photon direction $\bm{n}'$. 
Thus the power per unit solid angle in the scattered beam is
\begin{equation}\label{ch3:pscatt}
  {dP'\over d\Omega'}={3\sigma_{\rm T}\over8\pi} \,Q
\left\vert\bm\epsilon^
    {\ast}\cdot\bm\epsilon'\right\vert^2 \ .
\end{equation}
We may also write $Q\vert\bm
\epsilon^{\ast}\cdot\bm\epsilon'\vert^2=Q_{ij}
\epsilon^{\prime\ast}_{i} \epsilon^{\prime}_{j}$.  

Next consider a gas of electrons all at rest with number density
$n_e$. We can ignore the thermal motion of the electrons here
since the correction to the polarization due to a finite electron
temperature $T_e$ will be down by a factor of $\theta_e\equiv k_B
T_e/m_e c^2$. At cluster temperatures $\theta_e \approx 0.01$, so we
may assume stationary electrons. 
Assuming incoherent scattering, multiplying 
Eqn.~(\ref{ch3:pscatt}) by $n_e d\Omega'$ 
 converts scattered power per electron to the rate of change
of energy density in final polarization state $\bm\epsilon'$:
\begin{equation}\label{tscatt1}
  {d Q'_{ij}\over d t}\,\epsilon^{\prime\ast}_{i}
    \epsilon'_{j}={3\sigma_{\rm T}\over8\pi}n_e 
Q_{ij} \epsilon^{\prime\ast}_{i} \epsilon'_{j} \, d\Omega'\ ,
\end{equation}
where $Q'_{ij}$ is the polarization matrix of the scattered beam.
Using equation (\ref{ch1:def_fmunu}), 
setting $p=p'$ since we are working in the Thomson limit, 
and assuming the incident beam is monochromatic, we find
\begin{equation}\label{tscatt2}
  {d I'_{ij}\over dt}\,\epsilon^{\prime\ast}_{i}
    \epsilon'_{j} ={3\sigma_{\rm T}\over8\pi}n_e
\,\epsilon^{\prime\ast}_{i} \epsilon'_{j} \int d\Omega \;I_{ij}\ ,
\end{equation}
where $d\Omega$ is the
solid angle element associated with the incident photon direction
$\bm{n}$. Note that this equation is valid now for an arbitrary
incident radiation field.

We cannot now just remove the polarization factors and
conclude $d I^{\prime}_{ij}\propto \int d\Omega \,I_{ij}$ because the
polarization of the incoming wave does not lie in the same plane as 
the polarization of the scattered wave.  For a given outgoing
momentum $\bm p'$, the outgoing polarization is a linear
combination of two basis vectors $\bm\epsilon'_1$ and $\bm
\epsilon'_2$ (orthonormal and orthogonal to the photon momentum $\bm{p}'$).
Thus, $I_{ij}\epsilon^{\prime\ast}_{i} \epsilon'_{j}$
projects out of
the incoming matrix $I_{ij}$ only those components lying
in the $\bm\epsilon'_1$-$\bm\epsilon'_2$ plane.  This projection is
equivalent to first projecting $I_{ij}$ with $\bm
\epsilon'_1\otimes\bm \epsilon'_1+\bm \epsilon'_2\otimes\bm
\epsilon'_2$.  This is equivalent to projecting out the unphysical
components by acting with the matrix
\begin{equation}
P_{ij}(\bm{n'}) \equiv \delta_{ij} - n'_i n'_j \ .
\end{equation}
Projecting the final polarization vector with $P_{ij}(\bm{n}')$ does
not change it: $P_{ij}(\bm{n}')\epsilon'_{j}=\epsilon'_{i}$. 
It follows that $I_{ij}(\bm{n})\epsilon^{\prime\ast}_{i}
\epsilon_{j}=I_{kl}(\bm{n}) P_{ik}(\bm{n}') P_{jl}(\bm{n}')
\epsilon^{\prime\ast}_{i} \epsilon_{j}$.  Now it is safe to
remove the outgoing polarization vectors from
Eqn.~(\ref{tscatt2}). 

We conclude that, for any initial and final
polarizations,
\begin{eqnarray}\label{ch3:fscatt1}
  {d I'_{ij}(\bm{n}')\over dt} ={3\sigma_{\rm T}\over8\pi}n_e
P_{ik}(\bm{n}')\;P_{jl}(\bm{n}')
\int d\Omega \;I_{kl}(\bm{n}) \ .
\end{eqnarray}
This is the equation we need.
The projection tensors are easy to understand: the scattered
matrix is simply proportional to the incident matrix after
the unphysical polarization components (those proportional to $\bm
n'$) are eliminated.

If the integration time is sufficiently short, we may replace
$n_e \sigma_{\rm T}\,dt$ with the optical depth to Thomson scattering,
$\tau_{\rm C}$. Then we have for the polarization matrix of the
scattered radiation field:
\begin{eqnarray}
  I'_{ij}(\bm{n}') = {3\tau_{\rm C}\over8\pi}
P_{ik}(\bm{n}')\;P_{jl}(\bm{n}')
\int d\Omega \;I_{kl}(\bm{n}) \ .
\end{eqnarray}
If the incoming radiation field is assumed to be unpolarized, then
the incident matrix has the form of Eqn.~(\ref{appb:unpol})
and we may set $I_{kl}(\bm{n})=I(\bm{n})\;P_{kl}(\bm{n})/2$. Thus
\begin{eqnarray}
  I'_{ij}(\bm{n}') = {3\tau_{\rm C}\over16\pi}
P_{ik}(\bm{n}')\;P_{jl}(\bm{n}')
\int d\Omega \;I(\bm{n}) \;P_{kl}(\bm{n}) \ .
\end{eqnarray}
This is the form of the transfer equation used in
Eqn.~(\ref{ch5:poltensor}). 

Note that this is not the full transfer equation, since we have
ignored the effect of scattering out of the beam direction. 
But here we are only interested in the polarization generated by
scattering into the observation direction, and the loss of photons
from the beam only affects this at $O(\tau_{\rm C}^2)$.

Assuming a blackbody spectrum of incident photons, this transfer
equation also holds for the brightness temperature polarization matrix
(since Thomson scattering does not change the photon frequency, and
there is no Doppler shift because we have assumed the electrons are
stationary, thus the scattered radiation field also has a blackbody
spectrum). 
We may suppress the frequency dependence of 
the (brightness temperature) polarization matrix and associated Stokes
parameters of the scattered photon also, since there is no energy
transfer.

\section{Effective scale of the $z>0$ quadrupole  \label{testleff}}

\begin{figure}[tb] 
\centering 
\includegraphics[width=8cm]{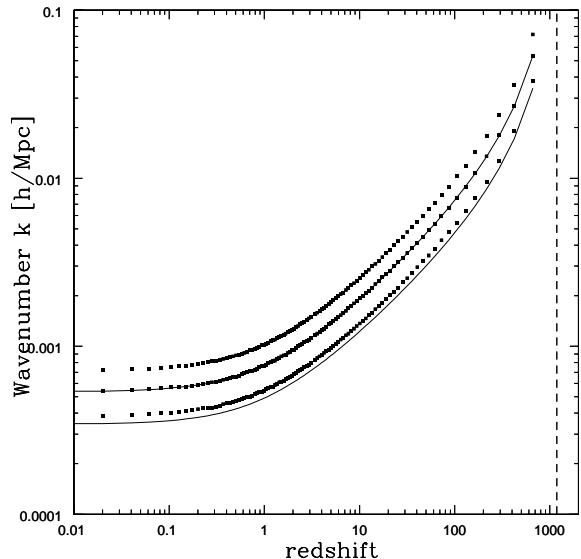}
\caption{
The lower, middle, and upper sets of points indicate the wavenumbers
at the 20th, 50th and 80th percentile respectively
of the window function for the $l=2$ harmonic, as a function of
redshift. This gives the range of wavenumbers 
contributing to the CMB quadrupole at a given redshift. 
The lower line shows the estimated wavenumber $k_{\rm{eff}}$
corresponding to the radius of the last scattering sphere. The upper
line shows $k'_{\rm{eff}}=3.12\;k_{\rm{eff}}$, which better
matches the median of the window function.
\label{ch4:kvzl2}
}
\end{figure}

The $l=2$ harmonic of the CMB anisotropy at redshift $z=0$
contains contributions from a broad range of wavenumbers of
order $c/H_0$.
At higher redshifts the CMB quadrupole probes the potential
on a smaller last scattering surface than at redshift $z=0$.
In \S\ref{ch5:sec_QUstat} we need to associate the quadrupole at
$z>0$ to an effective angular harmonic mode $l_{\rm{eff}}$ probing
roughly the same angular scales on the $z=0$ sky. This is most easily
done by finding the comoving wavenumber $k_{\rm{eff}}$
which best approximates the scale probed by the quadrupole on the
$z>0$ last scattering sphere, and setting
$l_{\rm{eff}}=2\;k_{\rm{eff}}/k_0$ where $k_0$ is an estimate of the
scale probe by the quadrupole at $z=0$. We take
$k_0=1/(\tau_0-\tau_r)$ (we have set $c=1$).

An approximation to the effective length scale probed by the CMB
quadrupole at a given redshift is given by the radius of the last
scattering sphere at that redshift, $r=(\tau_0-\tau_r)-\chi$
(where $\chi(z)$ is the comoving distance to redshift $z$).
With $k_{\rm{eff}}=1/r$ this gives
$l_{\rm{eff}}=2/(1-\chi/(\tau_0-\tau_r))$.

To check this approximation, 
there is a useful method \citep{2002PhRvD..66j3508T} which yields an
estimate of the range of wavenumbers $k$ which contribute to each $l$
mode. The procedure is to first find the following window function: 
\begin{equation} \label{ch4:pdfk}
\mathcal{P}(k|\tau,l) 
= \frac{(4\pi)^2\Delta^2_l(k,\tau) P_{\phi}(k)\;k^3}{C_l(\tau)},
\;\;\;\;\;\;\;\;
\int_{-\infty}^{\infty} \mathcal{P}(k|\tau,l) \;d \ln k = 1 \ ,
\end{equation}
where the transfer function is computed with some assumed
model. In the $l=2$ case the Sachs-Wolfe transfer function is sufficient
until we consider rather high redshifts.
The $k$ scale probed by this harmonic is then taken
to be the median $k_i$ of the window function.
The range of wavenumbers contributing to the $l$ harmonic at a
given redshift is indicated by 
the $20$th to the $80$th percentile of the window function
(which corresponds to full-width-half-maximum in the Gaussian case).
The $l=2$ case is shown by the points in Fig. \ref{ch4:kvzl2}. 
For comparison the estimate $k_{\rm{eff}}$ is shown by the lower line.
In fact we obtain a much better match to the
median window function estimate with the wavenumber
$k'_{\rm{eff}}=3.12 \;k_{\rm{eff}}$, as shown by the upper line. We
therefore use the following modified approximation for the effective angular
harmonic mode in the calculations in \S\ref{ch5:sec_QUstat}:
$l'_{\rm{eff}}=6.24/(1-\chi/(\tau_0-\tau_r))$.

\bibliographystyle{prsty}
\bibliography{main}

\end{document}